\documentclass[useAMS,usenatbib]{mn2e}
\usepackage{times}
\usepackage{natbib}
\usepackage{aas_macros}
\usepackage{amsmath}
\usepackage{amssymb}
\usepackage[pdftex]{graphicx}

\title[]{Simulations of the Galaxy Cluster CIZA J2242.8+5301 I: Thermal Model and Shock Properties}
\author[J. M. F. Donnert]{
	J. M. F. Donnert$^{2,1}$\thanks{donnert@ira.inaf.it}, A. M. Beck$^3$, K. Dolag$^3$, H. J. A. R\"ottgering$^4$    \\
$1$ INAF Istituto di Radioastronomia, via P. Gobetti 101, I-40129 Bologna, Italy\\
$2$ School of Physics and Astronomy, University of Minnesota, Minneapolis, MN 55455, USA\\
$3$ University Observatory Munich, Scheinerstr. 1, D-81679 Munich, Germany \\
$4$ Leiden Observatory, Leiden University, PO Box 9513, NL-2300 RA Leiden, the Netherlands
}

\begin{document}

\date{Accepted ???. Received ???; in original form ???}

\pagerange{\pageref{firstpage}--\pageref{lastpage}} \pubyear{2017}

\maketitle

\label{firstpage}

\begin{abstract}
	The giant radio relic in CIZA J2242.8+5301 is likely evidence of a Mpc sized shock in a massive merging galaxy cluster. However, the exact shock properties are still not clearly determined. In particular, the Mach number derived from the integrated radio spectrum exceeds the Mach number derived from the X-ray temperature jump by a factor of two. We present here a numerical study, aiming for a model that is consistent with the majority of observations of this galaxy cluster. We first show that in the northern shock upstream X-ray temperature and radio data are consistent with each other. We then derive progenitor masses for the system using standard density profiles, X-ray properties and the assumption of hydrostatic equilibrium. We find a class of models that is roughly consistent with weak lensing data, radio data and some of the X-ray data. Assuming a cool-core versus non-cool-core merger, we find a fiducial model with a total mass of $1.6 \times 10^{15}\,M_\odot$, a mass ratio of 1.76 and a Mach number that is consistent with estimates from the radio spectrum. We are not able to match X-ray derived Mach numbers, because even low mass models over-predict the X-ray derived shock speeds. We argue that deep X-ray observations of CIZA J2242.8+5301 will be able to test our model and potentially reconcile X-ray and radio derived Mach numbers in relics.
\end{abstract}


\section{Introduction}\label{sect.intro}

Merging galaxy clusters are among the most energetic events in the Universe. More than $10^{63} \,\mathrm{erg}$ of potential energy are released and dissipated in the inter-cluster-medium (ICM) on a time-scale of a Gyr \citep{2002ASSL..272....1S,2012ARA&A..50..353K}. Most of this energy is directed into heat through compression and shocks. These processes can be observed using X-ray satellites, which find signatures of shocks and cold fronts in many clusters \citep{1988xrec.book.....S,2007PhR...443....1M}. A small part of the potential energy has been argued to stir turbulence, amplify magnetic fields and accelerate relativistic particles \citep{2014IJMPD..2330007B}. These processes result in giant radio relics in many shocks and giant radio halos associated with the turbulent ICM of many merging clusters \citep{2004IJMPD..13.1549G,2012A&ARv..20...54F,2012SSRv..166..187B}. \par
A cluster prominently hosting all of these features is CIZA J2242.8+5301, the ''Sausage cluster''. Discovered by \citet{2010Sci...330..347V}, its nickname was coined after the northern relic of the cluster, which is evidence for a unique large shock propagating in the ICM. CIZA J2242.8+5301 itself is among the most well observed relic clusters in existence. Observational studies focus on the northern relic \citep{2013AA...555A.110S,2014MNRAS.441L..41S,2015AA...582A..87A,2016MNRAS.455.2402S}, the mass distribution \citep{2015PASJ...67..114O,2015ApJ...805..143D,2015ApJ...802...46J}, the structure of the thermal ICM \citep{2013PASJ...65...16A,2013MNRAS.429.2617O,2014MNRAS.440.3416O,2015AA...582A..87A}, the galaxy population \citep{2014MNRAS.438.1377S,2015MNRAS.450..630S,2015MNRAS.450..646S}. Theoretical studies address the problem of spectral steepening in the relic \citep{2015ApJ...809..186K,2014MNRAS.445.1213S,2016A&A...591A.142B,2016ApJ...823...13K} and magnetic field amplification in the shock \citep{2016MNRAS.462.2014D,2012MNRAS.423.2781I}. First simulations of the system have been presented early on by \citet{2011MNRAS.418..230V}. \par
Observations find different Mach numbers for the shocks and in the two relics. Mach numbers in the two relics derived from the total integrated radio spectrum are found to be in the range of 4.5 in the north and 2.8 in the south\footnote{We do not consider here Mach numbers derived from the radio spectral index profile. These are not robust, because spectral index maps can be resolution dependent and susceptible to projection effects} \citep{2013AA...555A.110S}. In contrast, the X-ray derived temperature jump across the shock gives values of 2.4 and 1.7 \citep{2015AA...582A..87A}. We  note that this difference in Mach number has significant implications for the shock energetics and thus models for cosmic-ray injection and magnetic field amplification, as the shock flux scales with the velocity to the third power.\par
Cosmological simulations generally find an abundance of internal cluster shocks with Mach numbers of two and above \citep[e.g.][]{2008MNRAS.391.1511H,2009MNRAS.395.1333V,2000ApJ...542..608M,2006MNRAS.367..113P,2008ApJ...689.1063S,2014ApJ...785..133H,2016MNRAS.461.4441S}. Mach numbers of more than four are found only in major mergers, so the radio derived Mach number cannot be excluded on these grounds. \citet{2015ApJ...812...49H} show that the projection of multiple shocks can potentially lead to an inconsistency in radio and X-ray derived Mach numbers in some configurations. However, the size and homogeneity of the relic in the north of CIZA J2242.8+5301 makes these configurations unlikely. \par
In this paper, we aim to find a numerical model for CIZA J2242.8+5301 that is consistent with the observations of the system. Significant new observational constrains emerged since the last numerical study of the cluster was attempted by \citet{2011MNRAS.418..230V}. Furthermore this study did not fully account for the DM dynamics of the system, which we model self-consistently. We can now also directly compare to the new weak lensing data. A consistent numerical picture of the system can provide clues to where to search for new observational evidence to reconcile the inconsistency in the Mach numbers in the shock and the relic.\par
We will use an idealized numerical model for merging galaxy clusters. Such models give us full control over the many merger parameters and allow us to efficiently explore the rather large parameter space \citep{1993LNP...421..267B,1993A&A...272..137S,1999ApJ...518..603R,2014ApJ...787..144L,2016arXiv160904121Z}. CIZA J2242.8+5301 is ideally suited for these kinds of simulations, as in contrast to other systems like the Toothbrush cluster, it is likely a simple two body merger. \par
This paper is structured as follows: We begin with a review and discussion of the current constrains for CIZA J2242.8+5301 from observations,  and present our approach  in section \ref{sect.sausage}. We outline our numerical model for spherically symmetric galaxy clusters and its implementation in section \ref{sect.model}. The results from the resulting simulations are elaborated in section \ref{sect.results} and discussed in section \ref{sect.discussion}. We draw our conclusions at the end in section \ref{sect.conclusions}.
We use a concordance cosmology with $h_{100} = 0.7$, $\Omega_\Lambda = 0.7$ and $\Omega_\mathrm{M} = 0.3$

\section{Models for CIZA J2242.8+5301} \label{sect.sausage}

\subsection{Observational Constraints}

\begin{figure}
	\centering
	\includegraphics[width=0.5\textwidth]{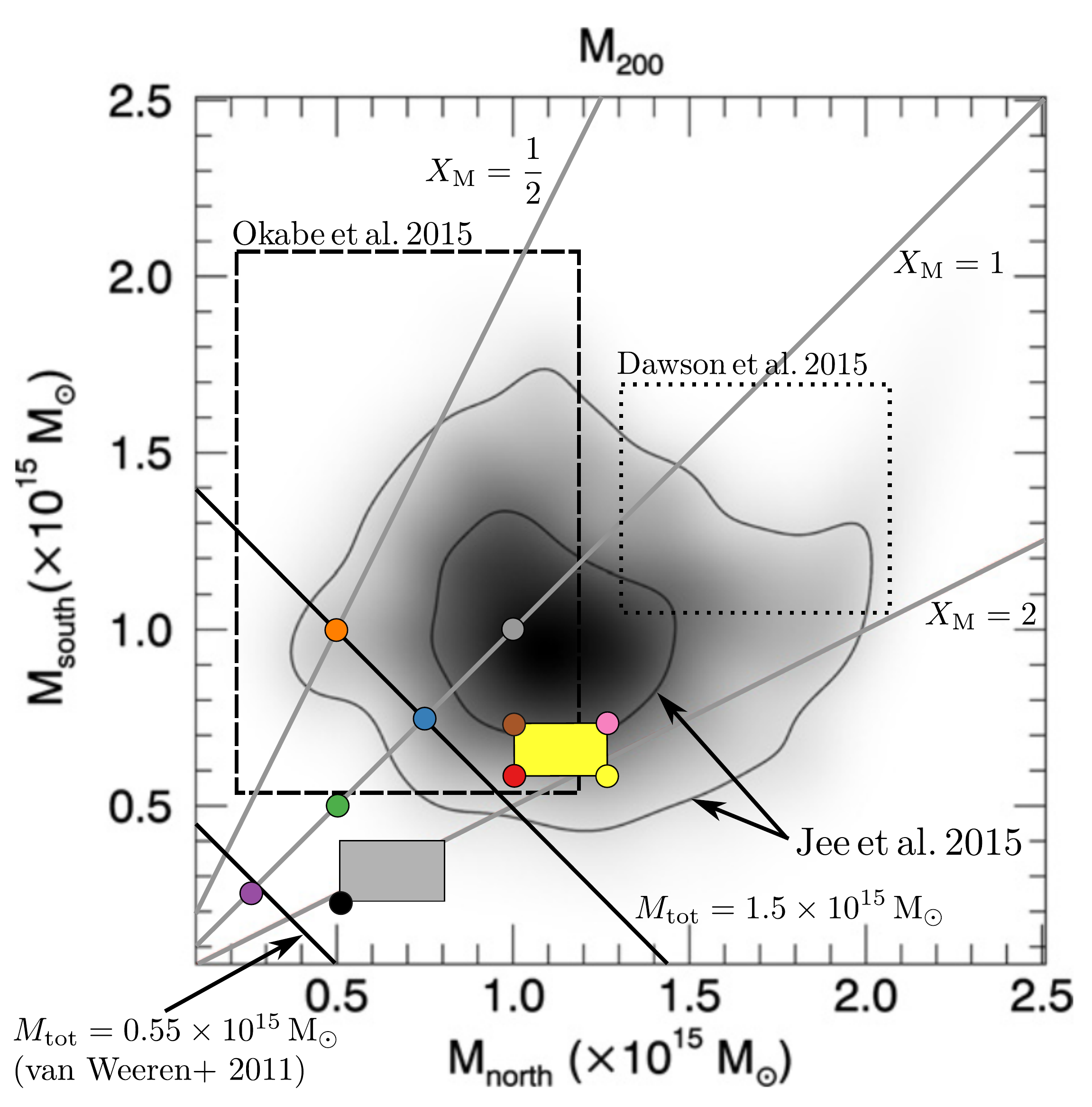}
	\caption{In contours, the two and one $\sigma$ confidence intervals of cluster masses from the weak lensing study (adapted from \citet{2015ApJ...802...46J}), weak lensing results from \citet{2015PASJ...67..114O} as black dashed box, strong lensing results from \citet{2015ApJ...805..143D} as black dotted box. Numerical models from this study as colored dots. We mark the lines with $M_\mathrm{tot} = 0.55 \times 10^{15} \,M_\odot$  \citep{2011MNRAS.418..230V} and $M_\mathrm{tot} = \, 1.5 \times 10^{15} \,M_\odot$   as black lines. Lines with  $X_\mathrm{M} = 0.5, 1, 2$  as grey lines. The plane of fiducial models with $r_\mathrm{cut} = 1.7 \, r_{200}$ is marked yellow, the plane with $r_\mathrm{cut} = 10 \, r_{200}$ in grey.} \label{fig.Lensing}
\end{figure}
 
Currently, CIZA J2242.8+5301 is among the best observed galaxy clusters in all of astronomy, especially in the radio band. Placed at a redshift of $z=0.188$ \citep{2015ApJ...805..143D}, the weak lensing study by \citet{2015ApJ...802...46J} finds sub-cluster masses\footnote{We give all masses relative to a top hat over density of $\Delta=200$.}  of $M_\mathrm{south} = 9.8^{+0.38}_{-0.25} \times 10^{15} \, M_\odot$ and  $M_\mathrm{north} = 1.1^{+0.37}_{-0.32} \times 10^{15} \, M_\odot$ with both mass peaks about $d_\mathrm{peak} = 1\pm0.15$ Mpc apart. Another analysis of the same data by \citet{2015PASJ...67..114O} yields $M_0 = 1.096^{+0.982}_{-0.567} \times 10^{15} \, M_\odot$, $M_1 = 0.551^{+0.639}_{-0.343}\times 10^{15} \, M_\odot$ and $d_\mathrm{peak} = 712 \,\mathrm{kpc}$, roughly consistent with the previous study. The probability distribution in the plane of subcluster masses from the former study are shown in figure \ref{fig.Lensing} (see \ref{sect.approach}), we add the constrains of the latter study as dashed box. These masses carry significant systematic uncertainties of up to a factor of two, because the mass distribution along the line-of-sight is not known and often assumed spherical, i.e. ignoring halo triaxiality (Hoekstra, priv. comm.).  \par

ROSAT finds a luminosity of $L_\mathrm{x} = 6.4 \times 10^{44}$ erg/s/Hz/cm$^2$ at $0.2 - 2.5$ keV \citep{2007ApJ...662..224K}. This brightness is roughly consistent with the value obtained by Chandra ($5.65\times 10^{44}\,\mathrm{erg}/\mathrm{s}$, 0.5 - 2.4 keV, Akamatsu \& Gu, priv.comm.). Suzaku observations by \citet{2015AA...582A..87A} find a temperature jump at the Northern Shock (NS) from $T_\mathrm{up,NS} = 2.7^{+0.9}_{-0.5}$ keV to $T_\mathrm{dw,NS} = 8.5^{+0.8}_{-0.6}$ keV, implying a Mach number of $M_\mathrm{NS} = 2.7$. At the Southern Shock (SS), $T_\mathrm{up,SS} = 5.1^{+1.5}_{-1.2}$ keV and $T_\mathrm{dw,SS} = 9^{+0.6}_{-0.6}$ keV, implying $M_\mathrm{SS} = 1.7$. \par

Radio observations find a radio halo and two radio relics at a distance of 2.8-3.2 Mpc, both exceeding 2 Mpc size at low frequencies  \citep[][ Hoang et al., submitted to MNRAS]{2016MNRAS.455.2402S,2010Sci...330..347V}. The northern relic (NR), the ''Sausage'',  is remarkably homogeneous and thin with a Mach number of $M_\mathrm{NR} \approx 4.6$. The irregular southern relic (SR) has $M_\mathrm{SR} \approx 2.7$. Both Mach numbers have been found from the total integrated spectrum \citep[][ Hoang et al. subm. to MNRAS]{2010Sci...330..347V,2016MNRAS.455.2402S}. In the simplest models, the spectral index profile of the NR is well fit by a downwind shock speed $v_\mathrm{dw,NR} \approx 1200$ km/s (\citet{2014MNRAS.445.1213S,2016MNRAS.462.2014D}). Comparing weak lensing and radio observations, we find that the NR has a distance of $d_\mathrm{NR} \approx 2\,\mathrm{Mpc}$ from the southern core, and the SR $d_\mathrm{SR} \approx 2 \, \mathrm{Mpc}$ for the northern core, a remarkably symmetric configuration. 

\subsection{Approach} \label{sect.approach}

\begin{figure}
	\centering
	\includegraphics[width=0.45\textwidth]{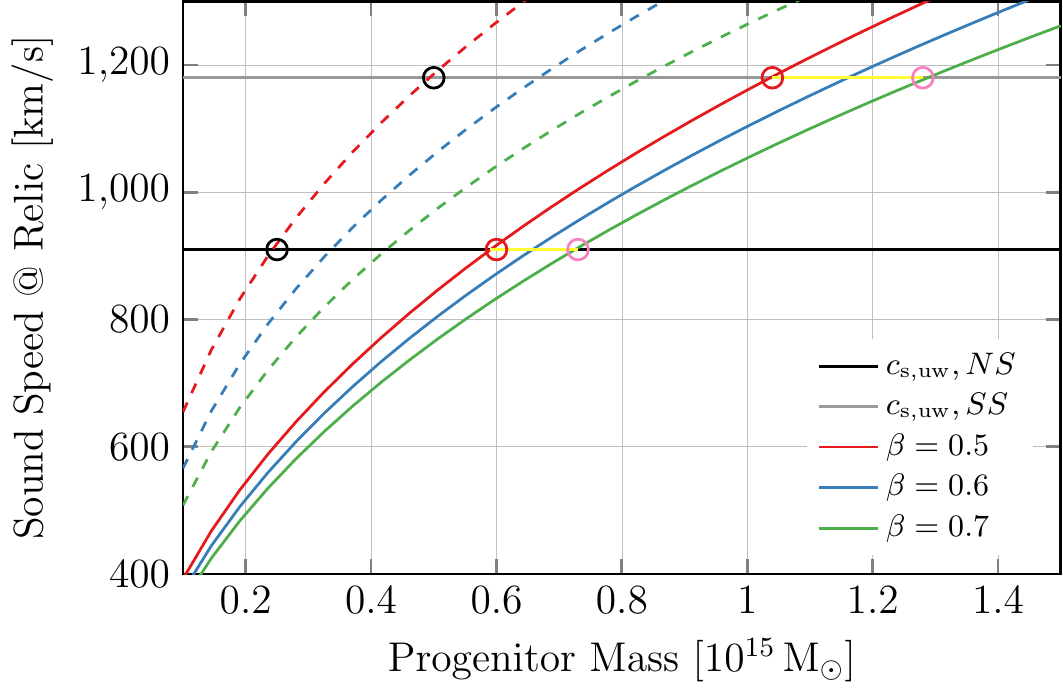}
	\caption{Sound speed at 2 Mpc over progenitor mass for $\beta = 0.5, 0.6, 0.7$ in red, blue and green respectively. We include the upstream sound speed inferred from observations as black and grey horizontal lines. We plot the shock speed for $r_\mathrm{cut} \rightarrow \infty$ as dashed lines. The yellow lines mark the mass range of fiducial models for radio scenario. We circle progenitor masses of the Black, Red and Pink models in their respective colors.} \label{fig.SoundSpeed}
\end{figure}

The X-ray morphology of the system is elongated along the merger axis. A simulated merger of two disturbed clusters leads to a displacement of both ICM's, which is not consistent with the narrow shape observed in CIZA J2242.8+5301. Hence, \emph{we adopt the working hypothesis that the northern progenitor was a cool core cluster}, which is breaking up during the ongoing merger with the southern progenitor  (we test this in appendix \ref{app.nonCC}). This naturally motivates the regular shape of the NR, which prohibits large local fluctuations of density, magnetic field and CR density in the upstream medium. To maintain this regular shape over its travel time of 0.5 to 1.0  Gyrs, the shock must have propagated through a cone free of larger fluctuations in sound speed/temperature and density. This is in contrast to the southern relic, that is rather irregular. Indeed, cosmological simulations suggest that this difference in shock structure is common in merging cluster \citep{2016MNRAS.461.4441S}. However, our idealized simulations are not able to faithfully reproduce the disturbed morphology of the southern shock. \par
To find a model for CIZA J2242.8+5301 that is consistent with the majority of observations, we first consider the northern shock to constrain its upstream medium and thus the progenitor of the southern sub-cluster. We can relate X-ray and radio observations with our model and its mass predictions with global cluster properties like weak lensing masses and X-ray brightness. 

\subsubsection{Radio Scenario}

In the NR, with a Mach number of $M = 4.6$, the Rankine-Hugoniot jump conditions give a compression factor of $\sigma_\mathrm{NR} = 3.5$. Combined with the downstream shock velocity inferred from the radio spectral index profile $v_\mathrm{dw,NR} \approx 1200 \,\mathrm{km}/\mathrm{s}$, this gives an upstream velocity of $v_\mathrm{up,NR} = 4200\,\mathrm{km}/\mathrm{s}$. As
\begin{align}
	M &= \frac{v_\mathrm{up}}{c_\mathrm{s}} \label{eq.M}
\end{align}
this corresponds to a sound speed of $c_\mathrm{s,NR} = 918 \,\mathrm{km}/\mathrm{s}$ ahead of the shock or a temperature of $T \approx 3.6 \times 10^{7}\,\mathrm{K} = 3.1\,\mathrm{keV}$. This is roughly consistent with the X-ray measurements for the upstream temperature of $2.9^{+0.9}_{-0.5}$ keV. The same estimates can be made for the southern shock and progenitor. However, without velocity constraints from the southern relic, which is highly irregular and possibly contaminated by outflows from close-by galaxies, so a spectral index profile does not lead to a clear downwind temperature. The upstream temperature from the X-rays (5 keV) gives $c_\mathrm{s,SR} \approx 1180\,\mathrm{km}/\mathrm{s}$, which with the $M_\mathrm{SR} = 2.7$ gives $v_\mathrm{up,SR} = 3305 \,\mathrm{km}/\mathrm{s}$. We call this parameter set of Mach numbers, upstream temperatures and shock velocities the \emph{radio scenario}.

\subsubsection{X-ray Scenario}

Above arguments can be turned around to predict the shock velocity from the measured X-ray temperature jump. The X-ray derived Mach number of $M_\mathrm{NS} = 2.7$ requires  $v_\mathrm{up,NS} = 2300\,\mathrm{km}/\mathrm{s}$ upstream of the NS. In the SS we infer from the X-ray temperatures:  $v_\mathrm{up,SS} = 2040\,\mathrm{km}/\mathrm{s}$. We call this parameter set, which is inconsistent with the radio parameter set,  the \emph{X-ray scenario}.  The difference in shock speed in the two scenarios ($v_\mathrm{up,NR} \approx 2 v_\mathrm{up,NS}$) illustrates the inconsistency of the X-ray and radio observations independently of cluster properties like temperature or mass.

\subsubsection{Mass Range}

As the ICM ahead of the shock has not been affected by the merger, we can use the upstream sound speed to choose the total mass of the progenitor given its $\beta$-model and the distance to the progenitor mass peak. In figure \ref{fig.SoundSpeed} we plot the sound speed at $d_\mathrm{Relic} = 2\,\mathrm{Mpc}$ over progenitor mass for a model with $\beta$ equal to 0.5, 0.6 and 0.7 in red, green and blue, respectively. We also overplot the observed upstream sound speed in the NR (SR) as black (gray) line. From the intersection of model sound speed with observed sound speed we find that both are consistent in a mass range from $M_0 = 0.59 - 0.73 \times 10^{15}\,M_\odot$ (yellow line), depending on the value of $\beta$. This mass range is also roughly consistent with the weak lensing value of the southern sub-cluster $M_\mathrm{south} = 0.98^{+0.38}_{-0.25} \times 10^{15}\,M_\odot$ (yellow area in fig. \ref{fig.Lensing}). \par
For the southern relic at a distance of 2 Mpc, our model predicts a mass of $M_1 = 1.04 - 1.28 \times 10^{15}\,M_\odot$ (figure \ref{fig.SoundSpeed}). This is again consistent with the weak lensing estimate for the northern sub-cluster of $M_\mathrm{north} = 1.1\pm 0.3 \times 10^{15}\,M_\odot$. We will attempt to use our simulations to predict the kinematics of mergers from these models, the expected shock morphology and velocity and the location of the DM mass peaks.  \par

\section{Cluster Model} \label{sect.model}

\begin{figure*}
	\centering
	\includegraphics[width=0.45\textwidth]{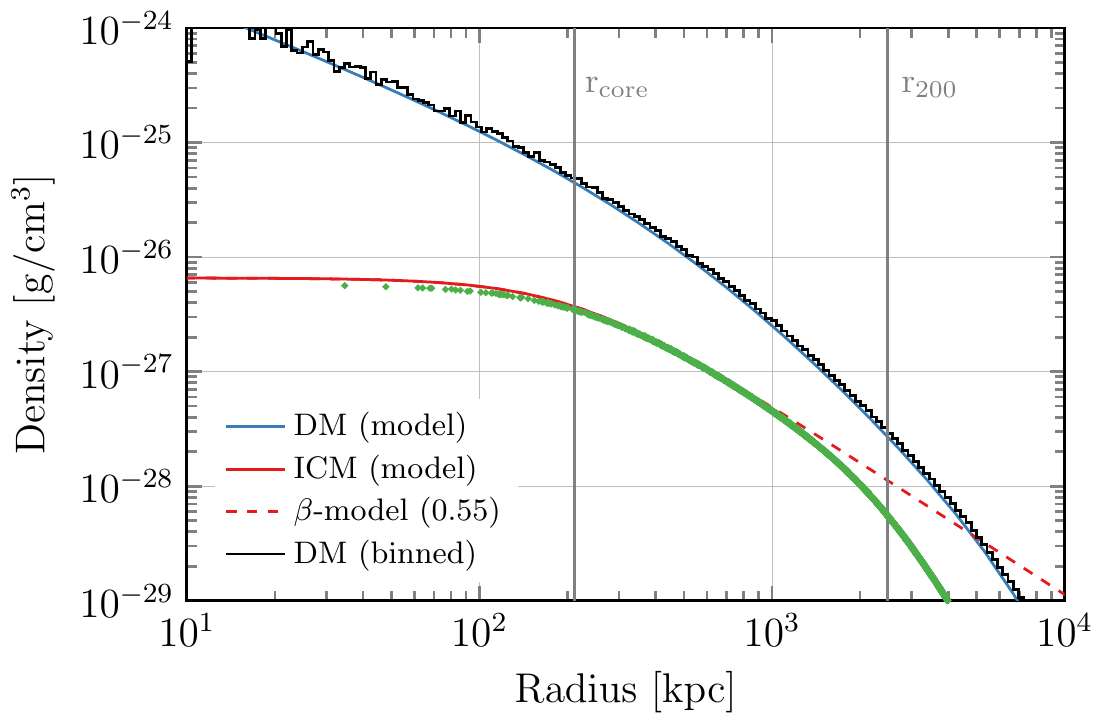}
	\includegraphics[width=0.45\textwidth]{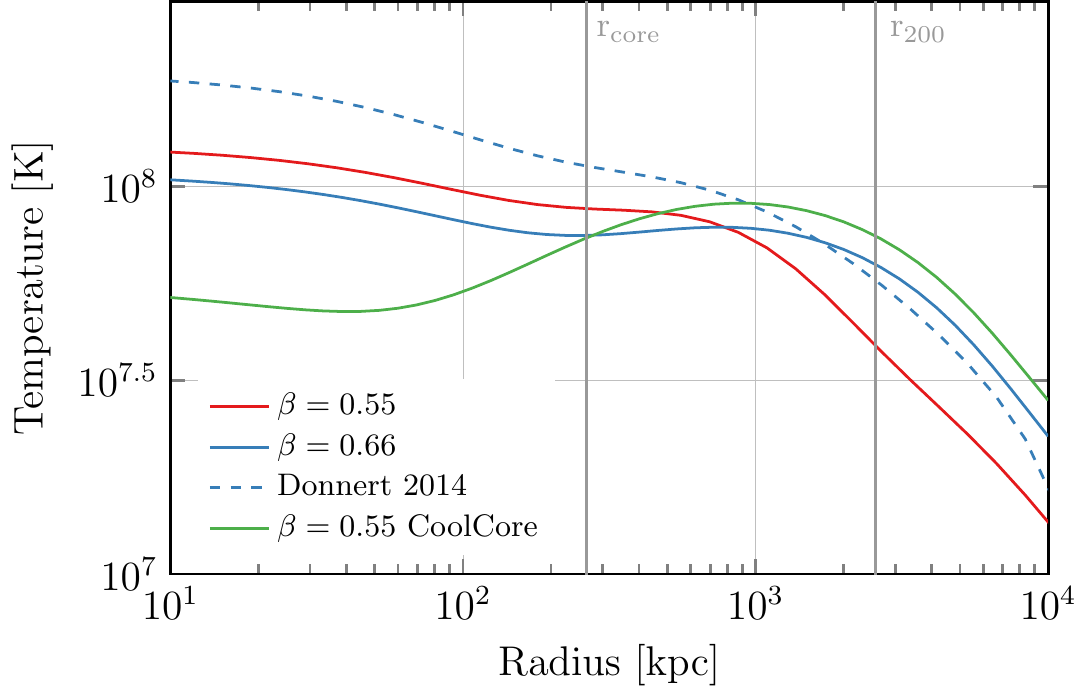}
	\includegraphics[width=0.45\textwidth]{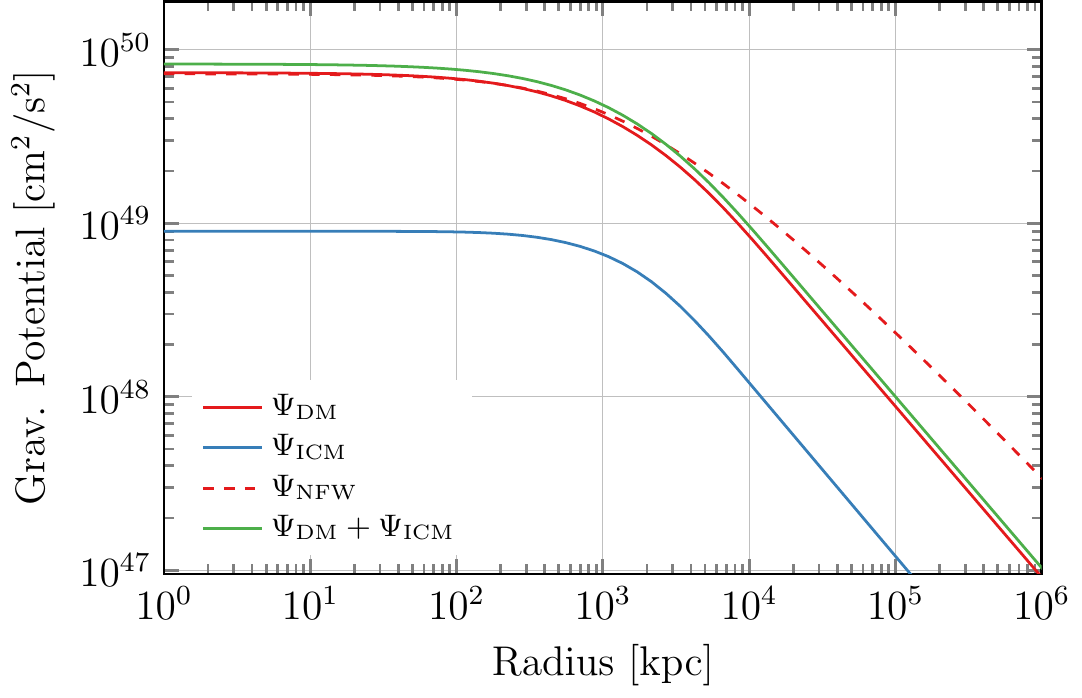}
	\includegraphics[width=0.45\textwidth]{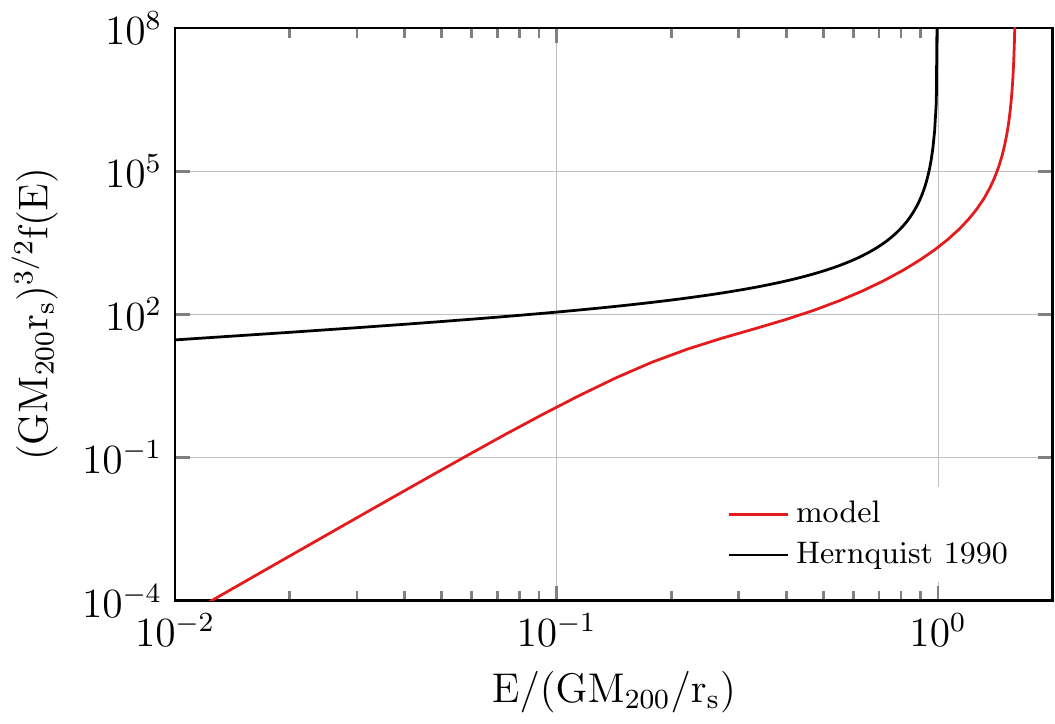}\\
	\caption{Initial conditions for a cluster with $M_{200} = 10^{15}\, M_\odot$ (top left to bottom right): Density profiles of gas (red), DM (blue), the standard $\beta$-model (red dashed), binned DM particles (black) and SPH density from a few thousand randomly selected WVT-regularized SPH particles (green dots); Top right: temperature profiles for values of $\beta = 0.66$ (blue) and 0.55 (red), a cool core with $\beta = 0.55$ (green) and the model from D14 (blue dashed); Bottom left: relative gravitational as generated from the DM density (red), the ICM (green), an NFW profile with the same DM mass (red dashed) and the sum of DM and ICM (green); Bottom right: distribution function of the DM particles (red) and from a Hernquist halo with the same mass (black). In the first two plots we mark $r_{200}$ and $r_\mathrm{core}$ with vertical grey lines.} \label{fig.IC_profiles}
\end{figure*}

We follow an approach similar to \citet[][ D14 hereafter]{2014MNRAS.438.1971D} to setup initial conditions for collisionless DM particles and SPH particles. We define the mass of the cluster as $M_{200}$ and then find $r_{200}$ as the radius where the average density of the cluster is $\Delta$ times the critical density at cluster redshift $z$ with $\Delta=200$. We assume the canonical baryon fraction ($b_\mathrm{f}$) of 17 percent in $r_{200}$ to find DM mass and ICM mass. A cluster is then completely defined by its DM and ICM density profiles and the assumption of hydrostatic equilibrium. We use an NFW-profile for the DM density and a beta-model for the ICM density \citep{1996ApJ...462..563N,1976A&A....49..137C}. Convergence demands that the models are cut-off at large radius: The NFW profile at the sampling radius $r_\mathrm{sample}$, which we set to  half the box size or $1.2 r_{200}$; The $\beta$-model is cut off at  $r_\mathrm{cut}$:
\begin{align}
	\rho_\mathrm{DM} &= \frac{\rho_{0,\mathrm{DM}}}{\frac{r}{r_\mathrm{s}} (1+\frac{r}{r_\mathrm{s}})}  \left(1+ \frac{r^3}{r^3_\mathrm{sample}}\right)^{-1} \\
	\rho_\mathrm{gas} &= \rho_{0,\mathrm{ICM}} \left(1 + \frac{r^2}{r_\mathrm{core}^2} \right)^{-\frac{3}{2}\beta} \left(1+\frac{r^3}{r_\mathrm{cut}^3}\right)^{-1} \label{eq.gas_density}
\end{align}

We then find the cumulative mass profiles, temperature profile and  relative gravitational potential ($\Psi = - \phi$) profiles from numerical integration using Gaussian quadrature from the GSL library \citep{contributors-gsl-gnu-2010}:
\begin{align}
	M(<r) &= 4 \pi  \int\limits_0^r \rho(t)\, t^2 \mathrm{d}t \\
	T(r) &= \frac{\mu m_\mathrm{p}}{k_\mathrm{B}} \frac{G}{\rho_\mathrm{gas}(r)}\int\limits^{R_\mathrm{max}}_r \frac{\rho_\mathrm{gas}(t)}{t^2} M_\mathrm{tot}(<t) \,\mathrm{d}t \\
	\Psi(r) &= G \int\limits^r_0\frac{M(<t)}{t^2} \,\mathrm{d}t,
\end{align}
where $G$ is Newtons constant, $k_\mathrm{B}$ is Boltzmanns constant, $m_\mathrm{p}$ is the proton mass and $\mu \approx 0.6$ is the mean molecular mass of the ICM plasma. We find the NFW scale radius $r_\mathrm{s}$ from the concentration parameter \citep{2008MNRAS.390L..64D} and set the core radius $r_\mathrm{core} = r_\mathrm{s}/3$ for non-cool core and $r_\mathrm{core} = r_\mathrm{s}/9$ for cool-core models, see \citet{2014MNRAS.438.1971D} and references therein.\par
To set the DM particle velocities we use rejection sampling \citep{1992nrfa.book.....P} from the particle distribution function $f(E)$. It is found from the combined gravitational potential of the gas and DM by numerically solving Eddingtons equation \citep{1916MNRAS..76..572E,2004ApJ...601...37K,2008gady.book.....B,2012MNRAS.425.1104B}:
\begin{align}
	f(E) &= \frac{1}{\sqrt{8}\pi^2}  \int\limits_{0}^{E} \frac{\mathrm{d}\Psi}{\sqrt{E - \psi}} \frac{\mathrm{d}^2\rho}{\mathrm{d}\Psi^2}\label{eq.Eddington}
\end{align}
by interpolating $\rho(\Psi)$ with a cubic spline and obtaining its second derivative directly from the spline. \par
We require an accurate SPH representation of the gas density distribution in our cluster model. Poisson sampling the ICM density field results in unacceptably large SPH sampling errors of $>20 \%$, severly affecting the hydrostatic equilibrium of the system. We use the technique of weighted Voronoi Tesselations to regularise the particle distribution and obtain smooth SPH densities \citep{2012arXiv1211.0525D}. We define a global density model as the maximum of the gas density of all clusters at a given position. A displacement can then be found from a neighbour search / SPH loop, which we implement using a Peano-Hilbert sorted oct-tree. The algorithm regularises the particle distribution in less than 20 iterations and the average SPH sampling error in density is reduced to less than five percent. Test simulation show that the resulting cluster models are numerically stable over several Gyrs. \par
We implemented our refined model into the C code mentioned in D14. The techniques employed in the IC generation are scalable enough to easily allow the generation of models with hundred million particles on modern SMP machines. 
In figure \ref{fig.IC_profiles} we show the model of a cluster with $M=10^{15}\,\mathrm{M}_\odot$:  In the first panel, we show the model density in red (ICM) and blue (DM) alongside the binned DM density (black line) and the SPH density on a random sample of 5000 SPH particles in green. We add the standard $\beta$-model as red dashed line, to show the effect of the cut-off in the ICM density. In the top right panel we show the ICM  temperature of the model and for a cluster with $\beta = 2/3$ as in \citet{2014MNRAS.438.1971D}. In these two panels we mark the core radius and $r_{200}$ as vertical grey lines. In the bottom left we show the relative gravitational potential (green), from the ICM (blue) and the DM (red). We add the potential from an NFW profile without cut-off as red dashed line. In the bottom right, we plot distribution function over relative energy $E = \Psi - v^2/2$ from the numerical solution of Eddingtons formula (eq. \ref{eq.Eddington}) using the combined DM \& gas generated potential, alongside the standard Hernquist solution (black \citet{1990ApJ...356..359H}).
 
\subsection{Setting the Cutoff Radius} \label{sect.rcut}

\begin{figure*}
	\centering
	\includegraphics[width=0.49\textwidth]{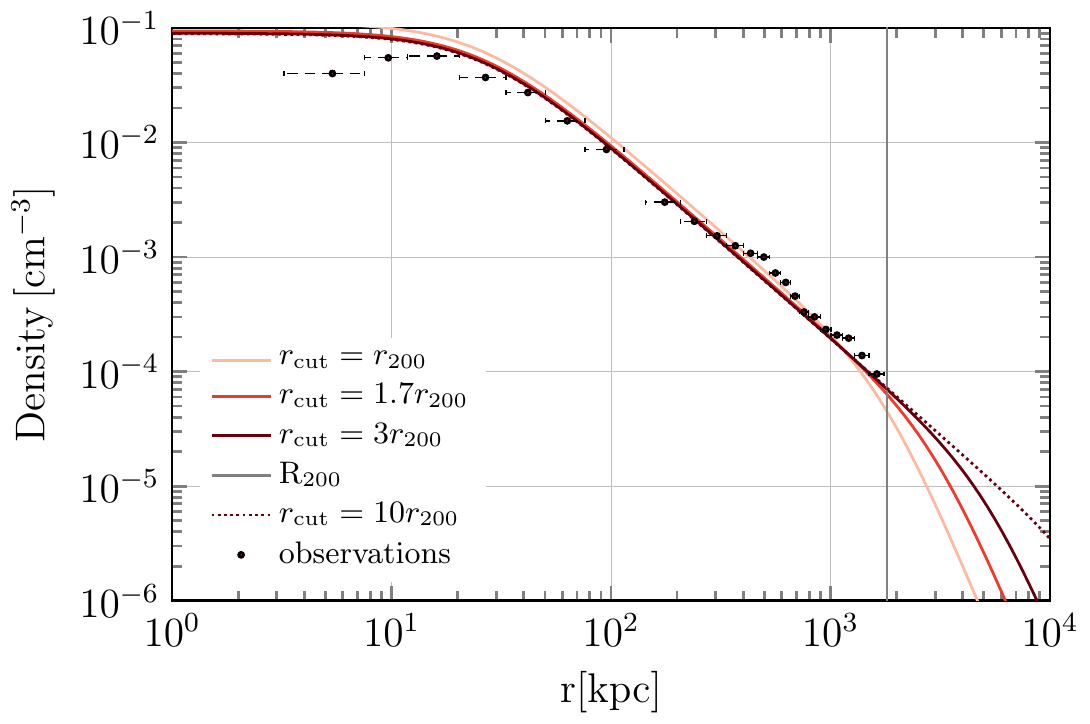}
	\includegraphics[width=0.46\textwidth]{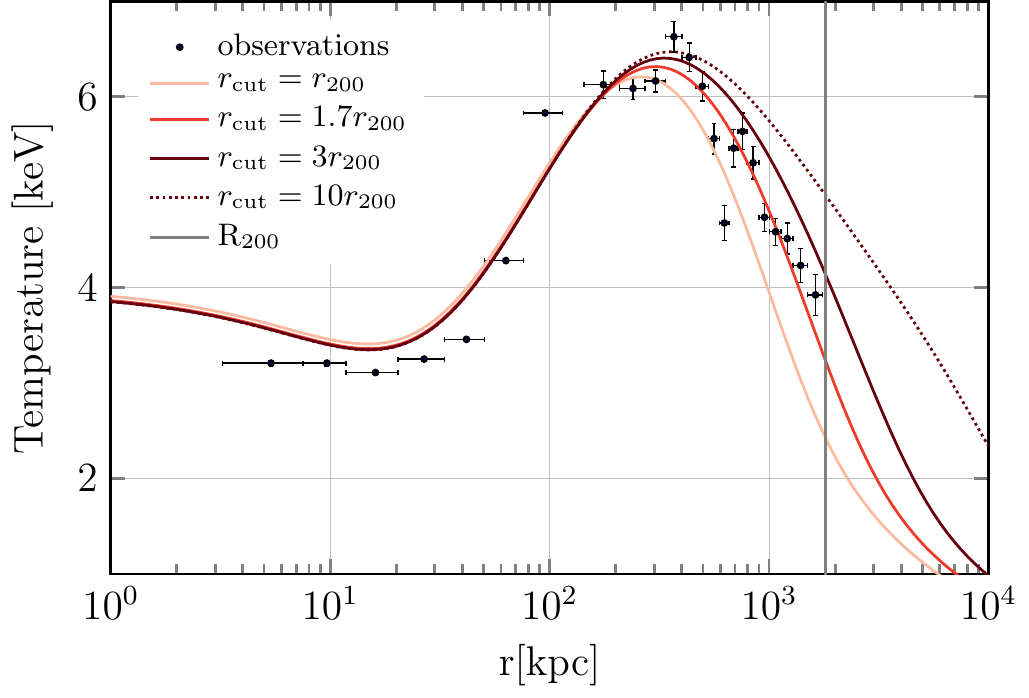}
	\caption{Left: Density profiles of three models for the Perseus cluster with $r_\mathrm{cut} = 1, 1.7, 3 \times r_{200}$ in light, bright and dark red, $r_\mathrm{cut} \rightarrow \infty$ as dotted line. $R_{200}$ as gray vertical line. Right: The same, but temperature profiles in keV. Observed profiles as black dots with error bars from \citet{2014MNRAS.437.3939U,2013MNRAS.435.3111Z}.} \label{fig.Trcut}
\end{figure*}

In figure \ref{fig.Trcut} we show the influence of $r_\mathrm{cut}$ (eq. \ref{eq.gas_density}) on density and temperature profiles fit to observations of the Perseus cluster. We show the deprojected profiles inferred from the X-rays as black dots with error bars \citep{2011Sci...331.1576S,2014MNRAS.437.3939U,2013MNRAS.435.3111Z}. In-line with previous fits, we find $M_{200} = 0.665\times10^{15} \,\mathrm{M}_\odot$, $r_{200} = 1810\,\mathrm{Mpc}$, $c_\mathrm{NFW} = 7.7$, $r_\mathrm{core} = 26 \,\mathrm{kpc}$, $\beta = 0.56$. We plot the temperature for $r_\mathrm{cut} = r_{200}$ (lighter red) and $r_\mathrm{cut} = 1.7 r_{200}$ (bright red) and $r_\mathrm{cut} = 3 r_{200}$ (dark red). We show $r_\mathrm{cut} \rightarrow \infty$ as dotted line. We find that the cut-off radius has significant impact on the temperature profile around the virial radius, with a best fit of $r_\mathrm{cut} = 1.7 r_{200}$. We  note that there is significant scatter in the data from the Perseus cluster. Depending on direction/arm of the observation, the data is consistent with $r_\mathrm{cut} = r_{200}$ and $r_\mathrm{cut} = 3 r_{200}$ \citep{2014MNRAS.437.3939U}. Thus $r_\mathrm{cut}$ is poorly constrained. Nonetheless, a model without cut-off (dotted line) is inconsistent with the Perseus temperature data beyond 500 kpc.

\section{Numerical Models}

\begin{table*}
	\centering
	\begin{tabular}{r|c|c|c|c|c|c|c|c|c|c|c}
		Name / Colour  & $M_\mathrm{0} $ & $M_\mathrm{1}$ & $M_\mathrm{tot}$ & $X_\mathrm{M}$ & $\beta_0, \beta_1$ & $L_\mathrm{x}$ & $M_\mathrm{NS}$ & $M_\mathrm{SS}$ & $v_\mathrm{NS}$ & $v_\mathrm{SS}$ & $t_\mathrm{travel}$ \\\hline\hline
		Grey  	& 1		& 1 	& 2 	& 1		& 0.6, 0.6	& 14 	& 3.9 & 4.5 &  4012 &  3897 & 600 \\
		Blue  	& 0.75	& 0.75 	& 1.5 	& 1		& 0.6, 0.6	& 10 	& 3.2 & 4.4 &  3215 &  3475 & 675\\
		Green  	& 0.5	& 0.5 	& 1 	& 1		& 0.6, 0.6	& 7.1 	& 3.8 & 4.7 &  2833 &  2779 & 775 \\
		Purple 	& 0.25	& 0.25 	& 0.5 	& 1		& 0.6, 0.6	& 3.7 	& 3.7 & 5.0 &  1973 &  1972 & 1000\\ \hline
		Red 	& 0.59	& 1.04 	& 1.63  & 1.76	& 0.5, 0.5	& 7 	& 2.4 & 3.2 &  2996 &  2971 & 675 \\ 
		Yellow 	& 0.59	& 1.28	& 1.88 	& 2.13	& 0.5, 0.7	& 11	& 3.6 & 3.5 &  3748 &  3408 & 650 \\
		Brown 	& 0.73	& 1.04	& 1.77 	& 1.43	& 0.7, 0.5	& 15	& 2.5 & 3.2 &  3323 &  2887 & 650 \\
		Pink 	& 0.73	& 1.28 	& 2.01 	& 1.75	& 0.7, 0.7	& 15 	& 3.8 & 3.4 &  3771 &  3137 & 675 \\ \hline
		Orange 	& 0.58	& 1.16	& 1.0 	& 0.5	& 0.6, 0.6	& 9.6	& 3.2 & 2.2 &  2076 &  1794 & 975 \\ 
		Black 	& 0.2	& 0.4	& 0.7 	& 2.5	& 0.5, 0.5	& 1.8	& 3.0 & 3.1 &  2150 &  1837 & 700 \\ 
		
	\end{tabular}
	\caption{Numerical models without initial velocity: Mass ($M_{200}$) of progenitors, total mass, mass ratio, $\beta$ parameter,  X-ray luminosity at observed state, upwind shock speeds and time between core passage and observed state. Names correspond to the colored dots in figure \ref{fig.Lensing}. Masses are in  $10^{15} \,M_\odot$, $L_\mathrm{x}$ is in $10^{44}$ erg/cm$^2$/s/Hz in the ROSAT band of 0.2 - 2.5 keV, shock speeds from upper 75th percentile of the Mach number distribution in km/s, time in Myr.} \label{tab.models}
\end{table*}

We model the northern progenitor as a cool core cluster by setting $r_\mathrm{core} = r_\mathrm{s}/9$.  The southern progenitor is modeled as disturbed motivated from the SR morphology with  $r_\mathrm{core} = r_\mathrm{s}/3$.   $\beta$ is generally degenerate with cluster mass, Baryon fraction and has a strong influence on X-ray brightness. Thus above considerations mark a region of acceptable progenitor masses in figure \ref{fig.Lensing}, which we mark yellow. Throughout this paper we name models after their color in this figure (\ref{fig.Lensing}). Models Red, Pink, Yellow and Brown sample the corners of this region. The difference between NS and SS Mach number already suggest that the mass ratio is not exactly one, regardless of the scenario. \par
We also consider models with $r_\mathrm{cut} \gg r_{200}$, to check the influence of the parameter. This allows us to predict the lowest progenitor masses corresponding to the lowest potential energies and thus shock speeds possible in our approach. We over-plot the sound speed at relic distance for a model with  $r_\mathrm{cut} \rightarrow \infty$ in figure \ref{fig.SoundSpeed} as dotted lines. Their intersection with the observed upstream temperatures gives mass of  $M_0 = 0.25 - 0.45 \times 10^{15}\,M_\odot$ for the northern progenitor and $M_1 = 0.5 - 0.81 \times 10^{15}\,M_\odot$ for the southern progenitor (grey area in figure \ref{fig.Lensing}). We call the most conservative model allowed in this range the Black model. \par
Furthermore, we sample the plane of progenitor masses at mass ratios of one (see figure \ref{fig.Lensing}), adopting a universal value of $\beta = 0.6$ (models Grey, Blue, Green, Purple). We add another model with inverted mass ratio (Orange).\par
We parameterize the in-fall velocity as a fraction of the zero energy/Kepler orbit $X_\mathrm{E}$, i.e. the velocity the clusters had if they were at rest at infinite distance (see also eq. \ref{eq.v0}). $X_\mathrm{E} = 1$ is the physical upper limit to the kinetic energy of the system. The lower limit is $X_\mathrm{E} = 0$, i.e. progenitors at rest when their distance of is the sum of their virial radii $r_{200)}$.\par
We focus on the X-ray and the shock properties of the simulations as predicted when the shocks are three Mpc apart. We first run all models with $X_\mathrm{E} = 0$, which minimizes Mach number in the shocks and gives a rough estimate on the X-ray luminosity of the system at the observed state. We then increase the in-fall kinetic energy to $X_\mathrm{E} = 0.5$. \par
We place the clusters at a distance so their virial radii touch and we set a small impact parameter of 50 kpc  to break the otherwise near perfect symmetry of the system (CIZA J2242.8+5301 is likely a head on collision). An overview of all models is provided in table \ref{tab.models}. \par

\section{Results} \label{sect.results}

We evolve all models with 10 million DM and 10 million SPH particles for 8 Gyrs on the Itasca cluster of the Minnesota Supercomputing Institute at the University of Minnesota, using the latest version of the Gadget-3 code, including magnetic fields and shock finding \citep{2009MNRAS.398.1678D,2016MNRAS.458.2080B,2016MNRAS.455.2110B}. We use Smac2 \citep{2014MNRAS.443.3564D} to compute projections from the simulation. X-ray brightnesses are given in the ROSAT band of 0.2 to 2.5 keV. We project the spectroscopic temperature, which better approximates the observed X-ray temperatures than the simulated temperature \citep{2004MNRAS.354...10M}.  We identify the simulated with the observed system when the shocks have a distance of ~3 Mpc. We find shock speeds and Mach numbers from the shock finder, where we use the mean of all particles with a Mach number above the 90th percentile of the distribution \citep{2016MNRAS.458.2080B}. \par

\begin{figure*}
	\centering
	\includegraphics[width=\textwidth]{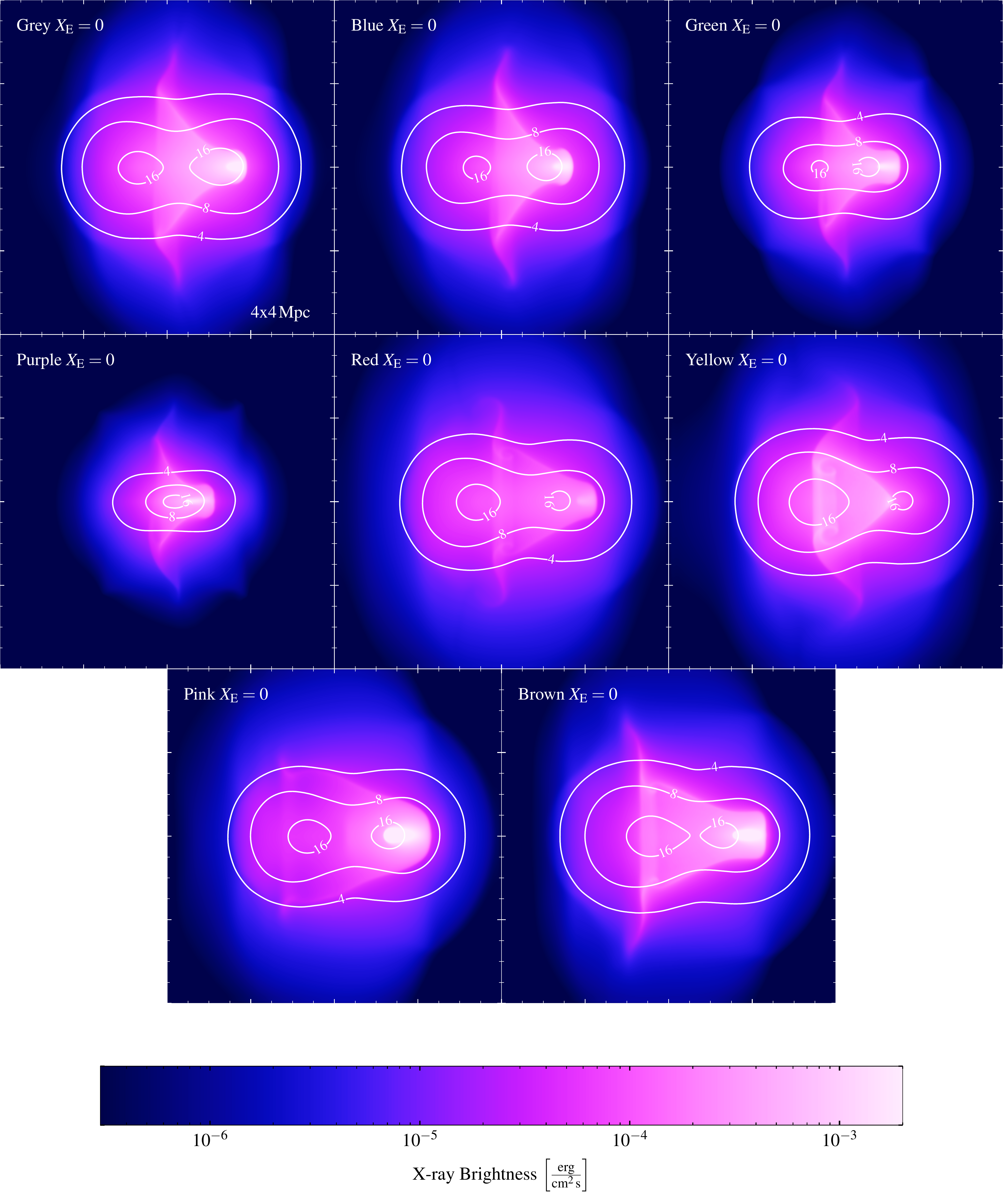}
	\caption{Projected X-ray emissivity of eight models with $X_\mathrm{E} = 0$ (Grey,Blue,Green,Purple,Red,Yellow,Pink,Brown). We overplot contours of the DM mass distribution in $10^{21} \,\mathrm{g}/\mathrm{cm}^2$.} \label{fig.Xslow}
\end{figure*}

\begin{figure*}
	\centering
	\includegraphics[width=\textwidth]{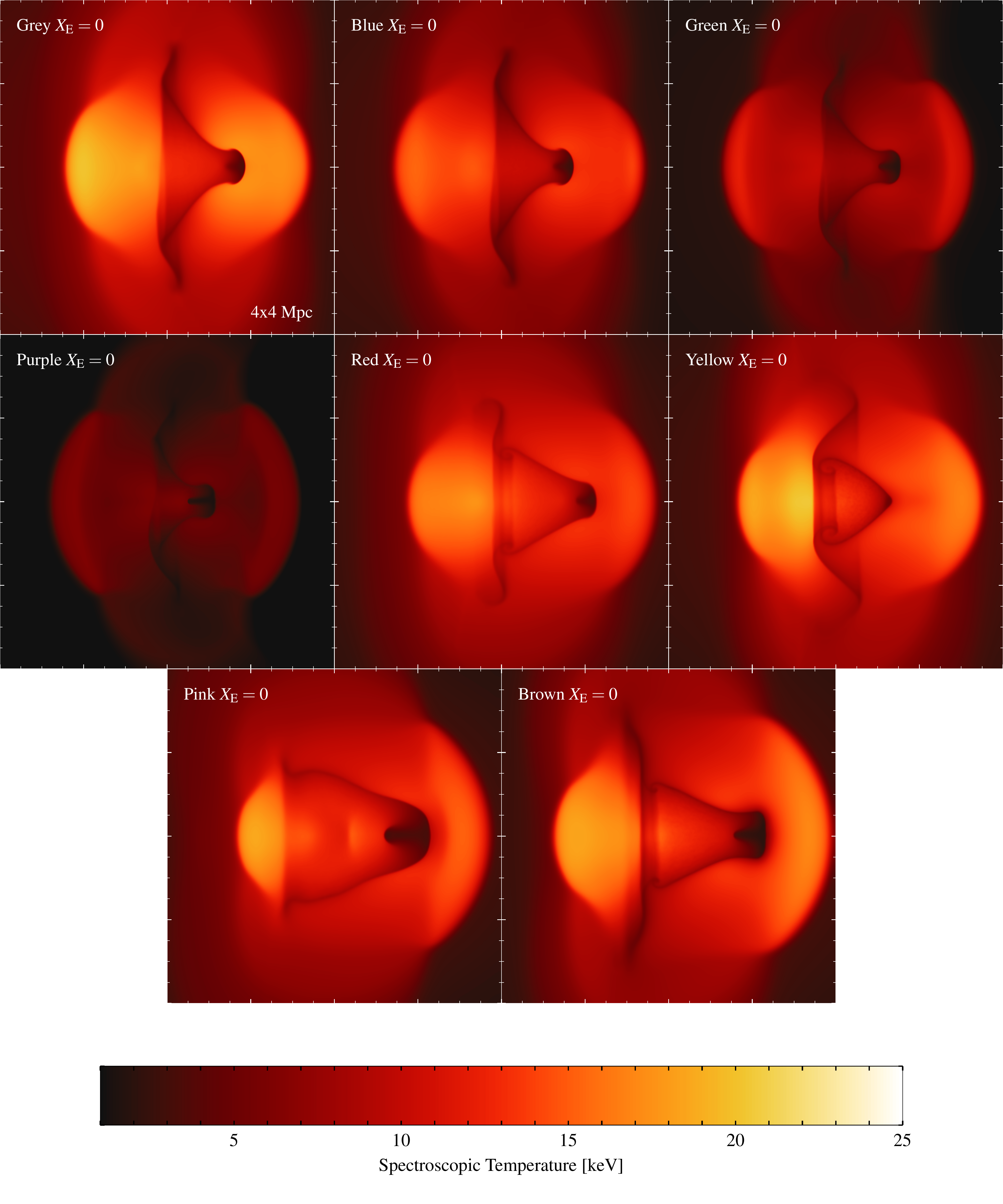}
	\caption{Projected Spectroscopic temperature of eight models with $X_\mathrm{E} = 0$ (Grey,Blue,Green,Purple,Red,Yellow,Pink,Brown).  } \label{fig.Tslow}
\end{figure*}

\subsection{Models with $X_\mathrm{M} = 1$}

\begin{figure}
	\centering
\includegraphics[width=0.45\textwidth]{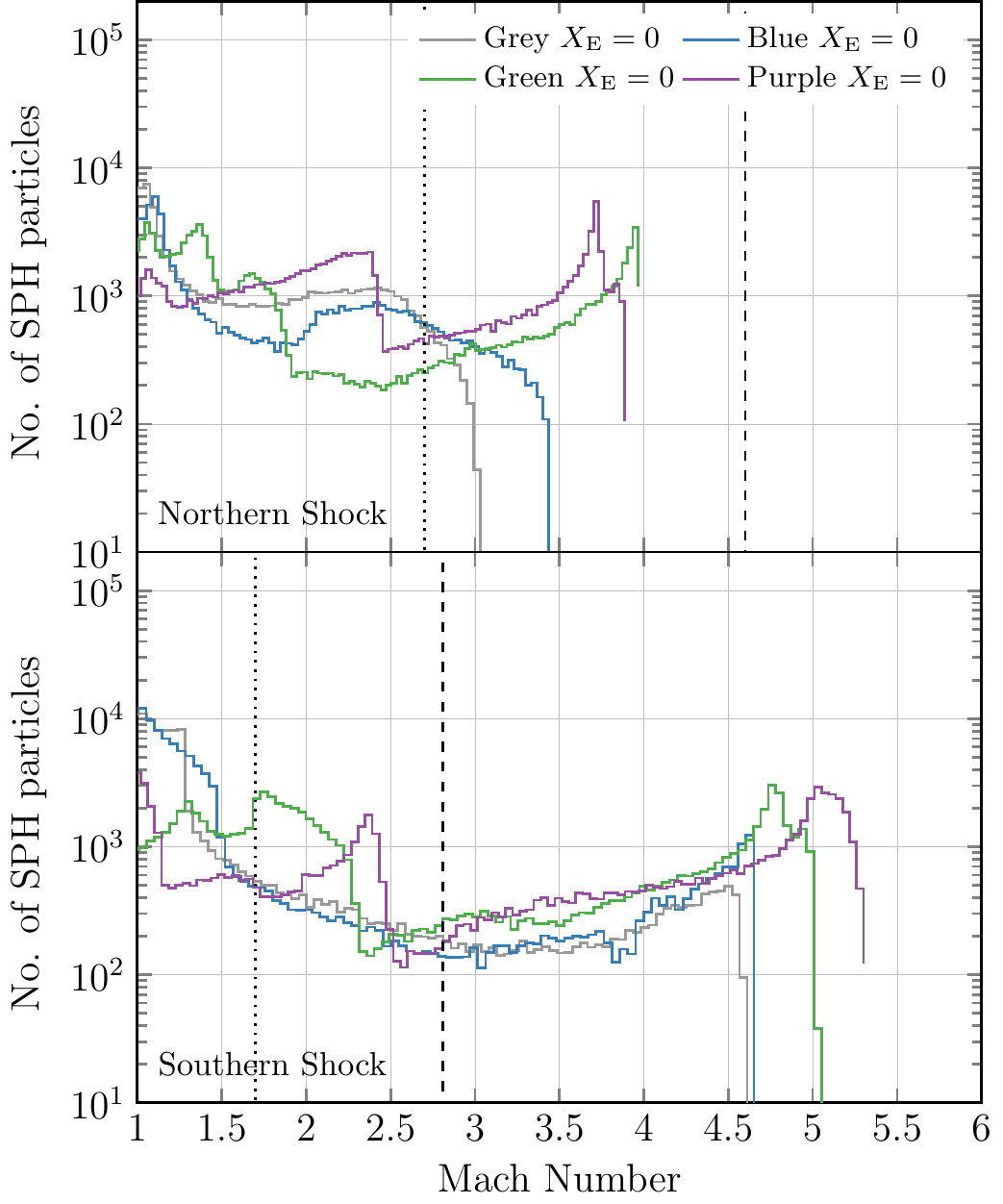}
	\caption{ Binned mach number distribution in the NS (top) and SS (bottom) from our mach finder for models with $X_{E} = 0$ and $X_\mathrm{M} = 1$. We also add the Mach number inferred from the radio (X-rays) in dashed  (dotted) vertical line.} \label{fig.mach_Xm=1}
\end{figure}

We begin with results from models with $X_\mathrm{M} = M_1/M_0 =  1$ and $X_\mathrm{E} = 0$. They represent a very conservative lower limit on the kinetic energy in the system. It is not likely that two clusters form at the distance of their virial radii. Thus we do not expect to find a well fitting model with these simulations. Furthermore a mass ratio of one minimizes the kinetic energy in the NS, because it matches the SS\footnote{This can easily be seen by considering a very large mass ratios. The smaller cluster will then not cause a large disturbance in the larger cluster.}. The goal is thus to obtain a lower limit on the  shock velocities and the X-ray brightness in the merger state for different masses. We give basic model parameters, X-ray luminosities, Mach numbers and $v_\mathrm{travel}$ at the observed state in table \ref{tab.models}. \par

In figure \ref{fig.Xslow}, panel 1-4 we show the projected X-ray luminosity of the four models. We find that the simulated X-ray emission has a triangular shape, less elongated than observed. The cool core has not broken up. Models with a total mass  of $M_\mathrm{tot} = 1-1.5 \times 10^{15}\, M_\odot$ are closest to the observed X-ray brightness. We also show the projected DM mass distribution in units of $10^{-21}\,\mathrm{g}/\mathrm{cm}^2$ as contours in figure \ref{fig.Xslow}. The distance between the DM mass peaks decreases with decreasing cluster mass. In all models but the heaviest one (Grey), the DM core has turned around, dragging ICM material with it.  For the lowest mass model (Purple), the two mass peaks have a separation of less than a few hundred kpc. This is not compatible with the weak lensing observations, that find a separation of about one Mpc with uncertainties of roughly $50 '' \approx 150 \,\mathrm{kpc}$ per core \citep{2015ApJ...802...46J,2015PASJ...67..114O}. \par
In figure \ref{fig.Tslow}, panel 1-4, we show the projected spectroscopic temperature of the models. All models show characteristic contact discontinuities, where the two ICMs pervade each other. This is likely a result of the idealised setup and not realistic. In the real system at least the southern progenitor has a disturbed morphology with bulk flows and density fluctuations that drive instabilities on multiple scales and facilitate mixing during the merger. This alters the X-ray morphology and temperature structure considerably (see section \ref{sect.sub}). Temperatures in the shocks are in the range of 10-15 keV in the cluster centre of the larger systems, which is in-line with the observations. \par 
All models show two symmetric shocks, whose size increase with decreasing cluster mass.  We find temperatures in all shocks ranging from 15 to 25 keV, with the highest temperatures in the NS of the lowest mass model (Purple). Shock speeds range from 2000 to 4000 km/s, increasing with cluster mass and similar in the NS and SS. Mach numbers range from three to five, with smaller Mach numbers in the NS. We speculate that the cool core drives the shock more efficiently than the disturbed progenitor.\par
In figure \ref{fig.mach_Xm=1} we plot the Mach number from our shock finder in the NS (top) and SS (bottom) for the four simulations, adding the observed Mach numbers from the shock and relic as dotted and dashed line, respectively. In the NS we find Mach numbers below 4, smaller than observed in the NR, and relative constant with cluster mass. In the SS the simulated Mach numbers of 4.5-5 exceed observed ones significantly. This is in-line with our expectations from section \ref{sect.approach}, where we argue for mass ratios above one, resulting in larger $M_1$ and thus higher temperatures and lower shock speed ahead of the SS. Thus models with $X_\mathrm{M} \le 1$ are disfavoured by X-ray and radio data alike. Our exploratory simulations suggest that the observed system has a mass ratio of above one, a total mass of $1-1.5 \times 10^{15}\,M_\mathrm{sol}$ and $X_\mathrm{E} > 0$.

\subsection{Models with $X_\mathrm{M} > 1$}

\begin{figure}
	\centering
\includegraphics[width=0.45\textwidth]{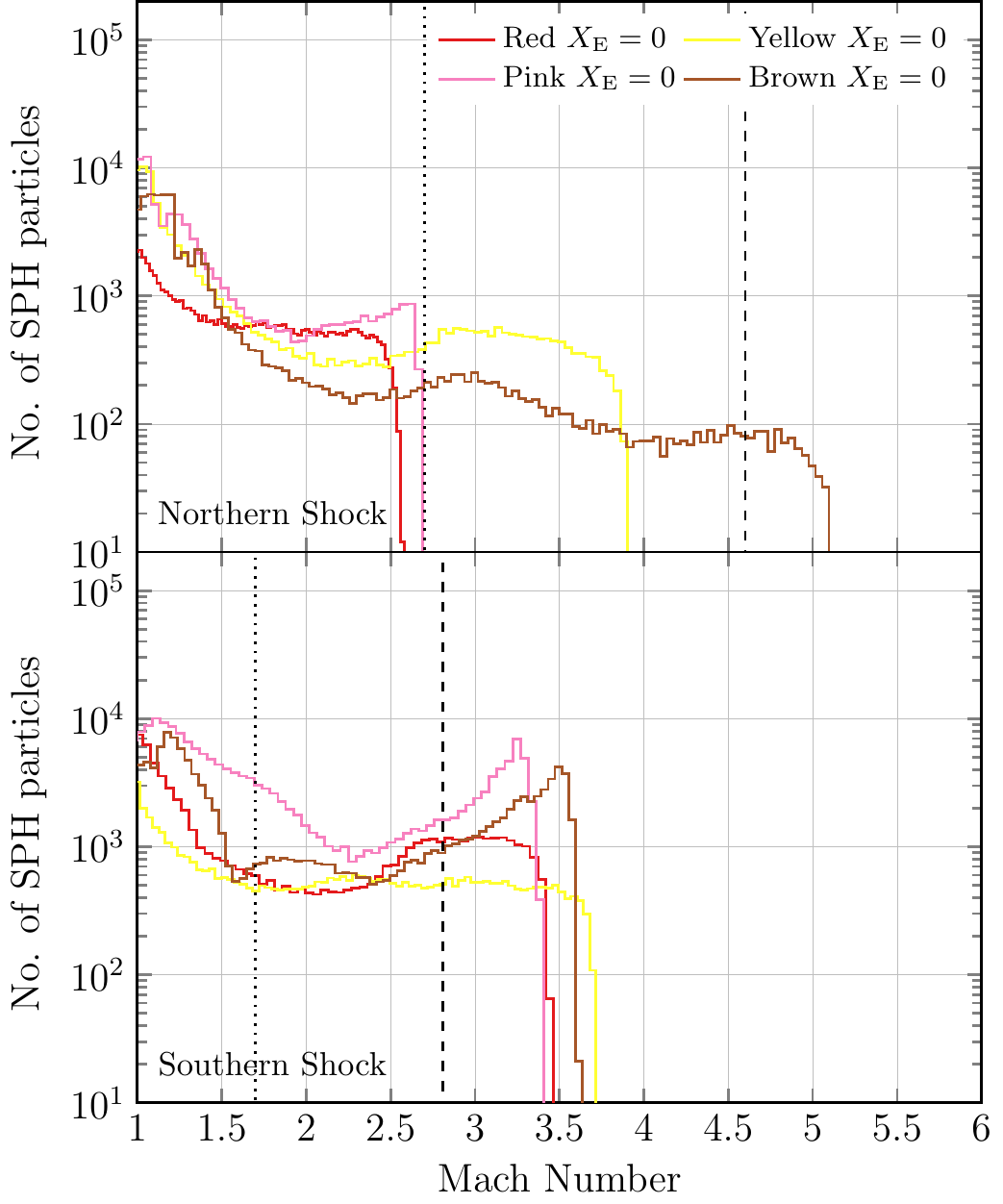}
	\caption{ Binned mach number distribution in the NS (solid) and SS (dashed) from our mach finder for models with $X_{E} = 0$ and $X_\mathrm{M} = 1$.} \label{fig.mach_slow}
\end{figure}

We now consider models with mass ratios above one (Red, Pink, Yellow, Brown), also shown in table \ref{tab.models}. As show above, every mass ratio implies a different values for the slope of the ICM profile ($\beta$). We find a strong dependence of the X-ray luminosity on this parameter. Only the Red model with $\beta_0 = \beta_1 = 0.5$ is close to the observed one, the other steeper models (Red, Pink, Yellow) are too bright (assuming a baryon fraction of $17\%$). As shown in figure \ref{fig.Xslow}, panel 5-8, the shape of the X-ray emission remains triangular, with subtle differences among models.  In all but the Yellow model the DM core of the northern progenitor has turned around. DM core separation is roughly one Mpc in all models. Temperature maps (figure \ref{fig.Tslow}, panel 5-8) show similar temperatures in the center of the cluster as before.\par
Again two shocks are clearly visible in the temperature maps, with post shock temperatures easily reaching 25 keV in the NS. However only the SS is fully developed. Mach numbers of these shocks range from 2.4 to 3.9 in the NS and 3.2 to 3.5 in the SS (table \ref{tab.models}) with shock velocities around 3000 km/s. The mach number distribution is shown in figure \ref{fig.mach_slow}, with quite similar distributions in the SS. However, in the NS the brown and yellow model show a tail in the distributions up to large Mach number of four and 5, respectively. This suggests that these two shocks are about to enter a region with lower temperatures and sound speeds, which boosts the mach numbers significantly.\par
We conclude that none of the models are consistent neither the radio nor the X-ray scenario.

\subsection{Models with Initial Velocity} \label{sect.fiducial}

\begin{figure*}
	\centering
	\includegraphics[width=\textwidth]{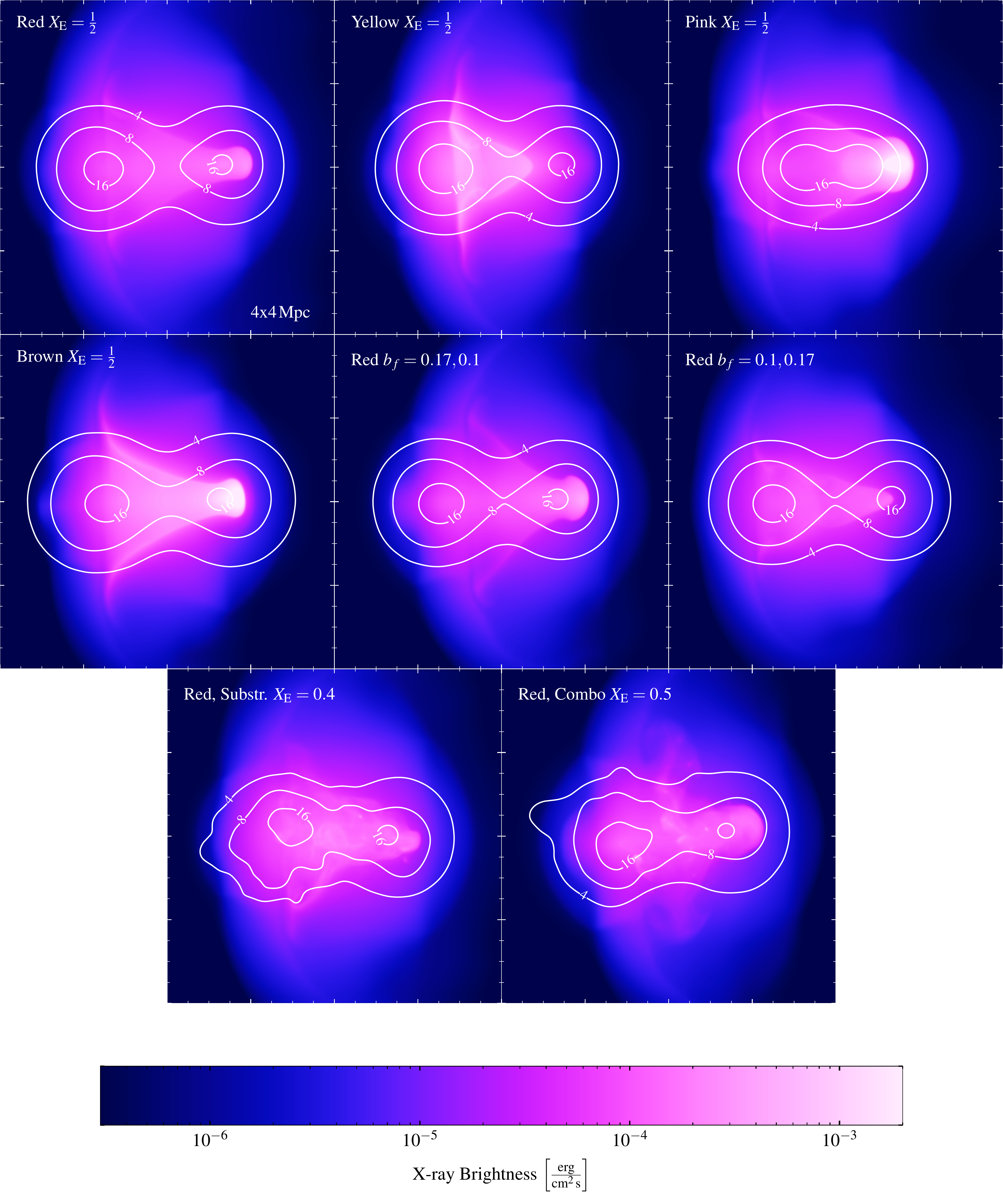}
	\caption{Projected X-ray emissivity of eight models with initial velocity: standard Red,Yellow,Pink,Brown and Red with reduced Baryon fraction, including substructure and including both. We overplot contours of the DM mass distribution in $10^{21} \,\mathrm{g}/\mathrm{cm}^2$.} \label{fig.Xfast}
\end{figure*}

\begin{figure*}
	\centering
	\includegraphics[width=\textwidth]{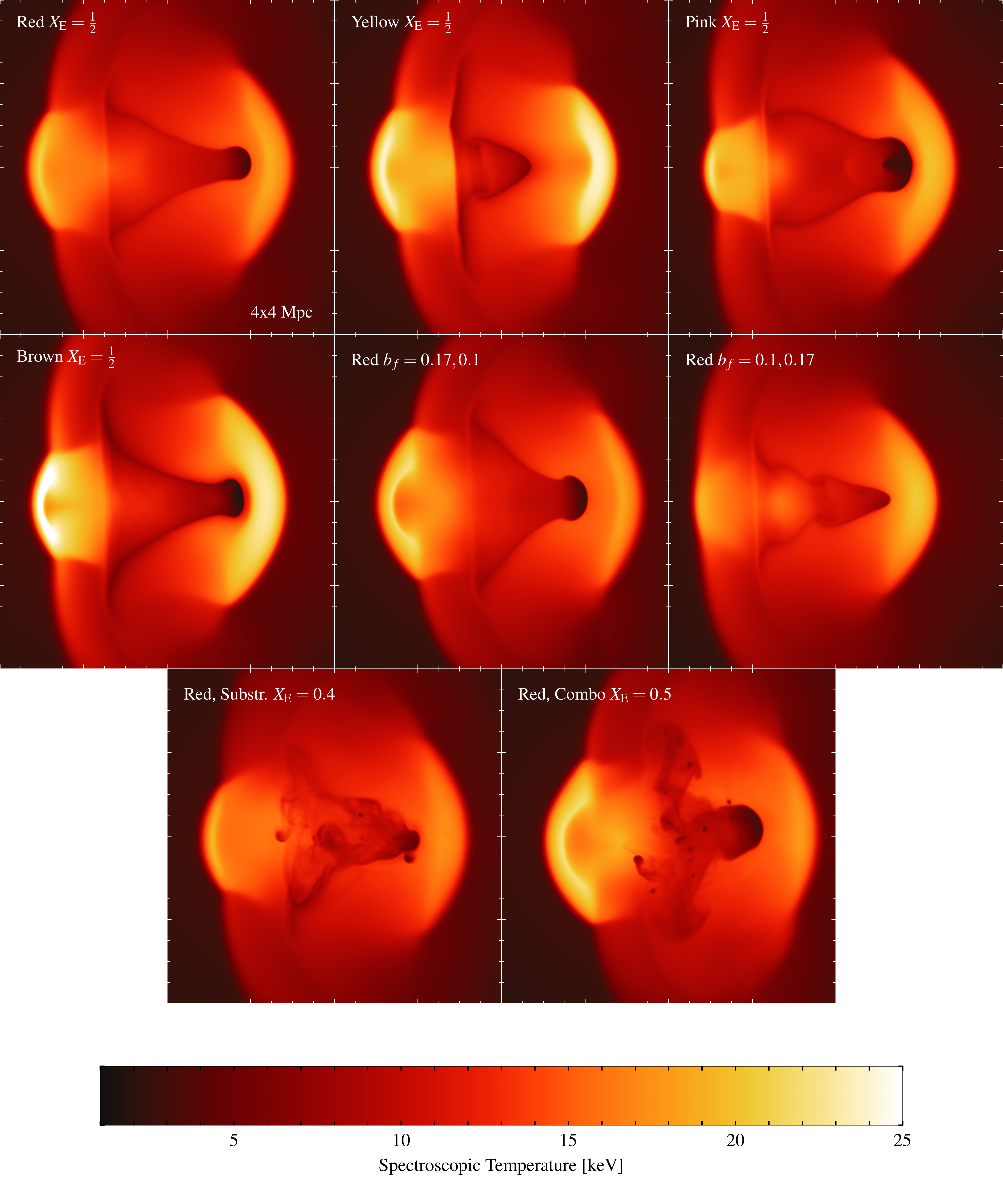}
	\caption{Projected Spectroscopic temperature of  eight models with initial velocity: standard Red,Yellow,Pink,Brown and Red with reduced Baryon fraction, including substructure (Red, Substr.) and including both (Red, Combo).  } \label{fig.Tfast}
\end{figure*}

\begin{table*}
	\centering
	\begin{tabular}{r|c|c|c|c|c|c|c|c|c|c}
		Name  & $b_{f,0}$& $b_{f,1}$ & $L_\mathrm{x}$ & $v_{0,0}$ & $v_{0,1}$ & $M_\mathrm{NS}$ & $M_\mathrm{SS}$ & $v_\mathrm{NS}$  & $v_\mathrm{SS}$ & $t_\mathrm{travel}$  \\\hline\hline
		red 	& 0.17& 0.17&  6.3 	& 664 &  -521 & 4.8 & 3.3 &  4367 &  3324 & 625\\ 
		yellow 	& 0.17& 0.17&  11 	& 746 &  -548 & 4.9 & 3.5 &  4647 &  3951 & 500 \\
		brown 	& 0.17& 0.17&  16 	& 628 &  -534 & 5.5 & 3.3 &  5150 &  3352 & 575 \\
		pink 	& 0.17& 0.17&  14 	& 710 &  -558  & 5.6 & 3.4 &  5262 &  3774 & 500\\ \hline
		Red, sub   & 0.17& 0.17  & 7 	& 531 &  -333 & 4.6 & 3.3 &  4221 &  3399 & 675  \\ 
		Red, combo & 0.17& 0.1 & 5.2 	& 664 &  -521 & 4.4 & 3.4 &  4081 &  3251 & 575 \\ \hline
		black 	& 0.17& 0.17&  2 	& 565 &  -395 & 3.0 & 4.6 &  2253 &  3259 & 600\\
		orange 	& 0.17& 0.17&  7.8 	& 338 &  -507 & 5.1 & 2.7 &  3165 &  2040 & 775 \\
	\end{tabular}
	\caption{Model name, Baryon Fraction, X-ray luminosity at observed state in $10^{44}$ erg/cm$^2$/s/Hz, initial merger velocities in km/s, Mach number of NS and SS in km/s,  shock speeds of NS and SS for numerical models with $X_\mathrm{E} = \frac{1}{2}$.} \label{tab.models_fast}
\end{table*}

\begin{figure}
	\centering
	\includegraphics[width=0.45\textwidth]{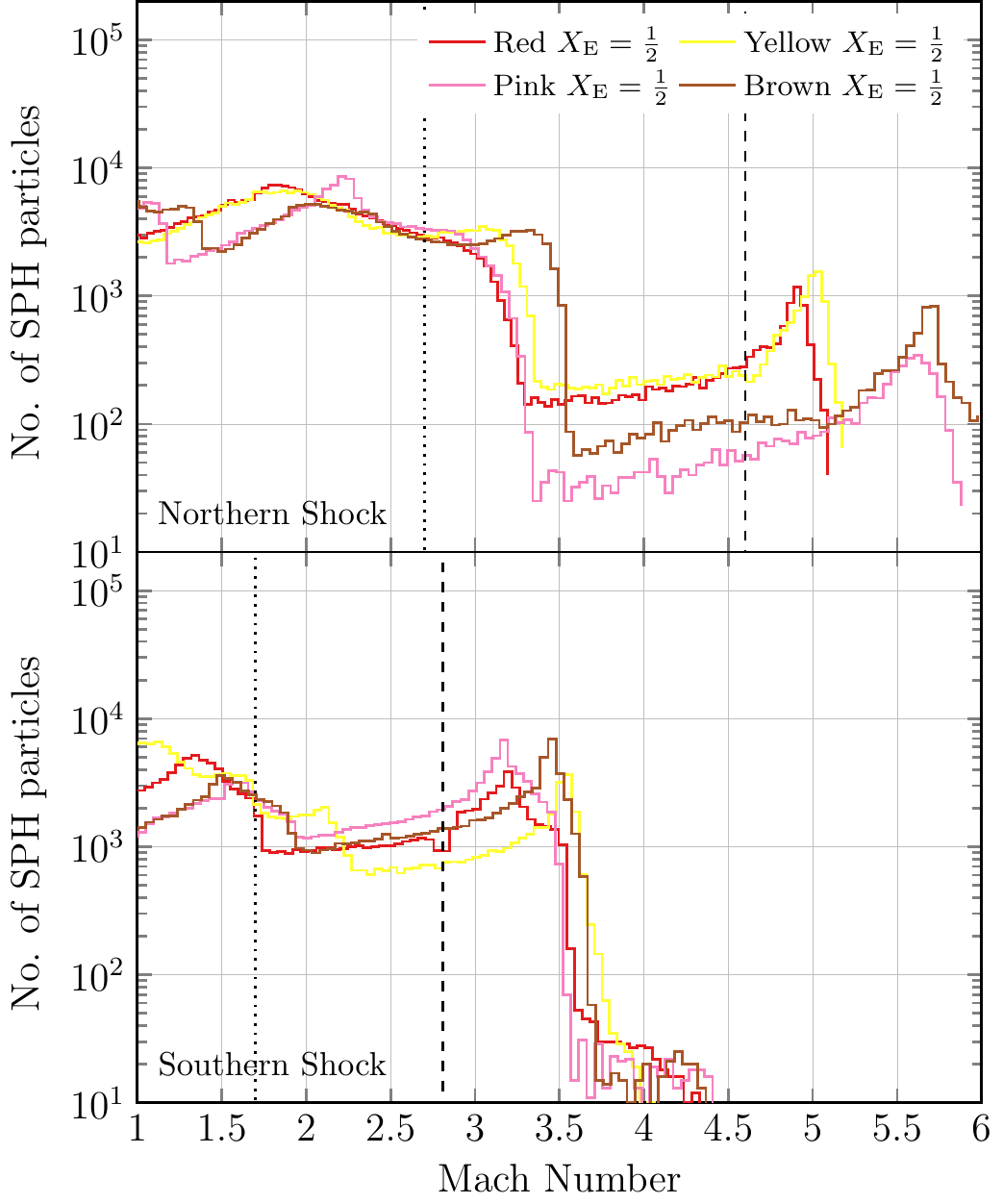}
	\caption{ Binned mach number distribution in the NS (top) and SS (bottom) from our mach finder for the Black model with $X_\mathrm{E} = 0$ and $X_\mathrm{E} = 0.5$.} \label{fig.mach_fast}
\end{figure}

We re-simulate models with mass ratios above one (Red, Yellow, Brown, Pink) with initial velocity, given by a zero energy orbit of $X_\mathrm{E} = 0.5$. The initial velocity of the system is shown alongside X-ray brightness, Mach numbers and shock velocities in table \ref{tab.models_fast}. We show projections of X-ray brightness and overlay them with DM density contours in figure \ref{fig.Xfast}, panel 1-4, and Mach number distributions in figure \ref{fig.mach_fast}. \par
The total X-ray luminosity remains roughly unchanged with respect to the slow versions of the models, with the Red model reproducing the observed X-ray luminosity. The shape of the X-rays is now elongated along the merger axis, all models show a mass peak separation of 1.2 - 0.7 Mpc, consistent with observations. Temperatures in the cluster center are lower than in the slow models, around 10 keV consistent with observations. \par
Temperature projections in figure \ref{fig.Tfast}, panel 1-4 show two clear shocks in all models, with the NS significantly smaller than the SS. Temperatures in the NS reach 25 keV in all models, while it varies in the SS from 20 keV (Red, Pink) to above 25 keV (Yellow). Mach numbers in the NS are above 4.5, and low with around 3.5 in the SS. Shock velocities range from 4300 to 5300 km/s in the NS and 3300 to 4000 km/s in the SS.  Mach number distributions (figure \ref{fig.mach_fast}) confirm these values, with sharply peaked distribution at Mach numbers above 4.5 (NS) and 3.3 (SS). We conclude that the Red model is roughly consistent in the radio scenario in Mach number and shock speed. However the elongated size of the NS is too small. 

\subsection{Two Models with Different Baryon Fractions} \label{sect.bf}

\begin{figure}
	\centering
\includegraphics[width=0.45\textwidth]{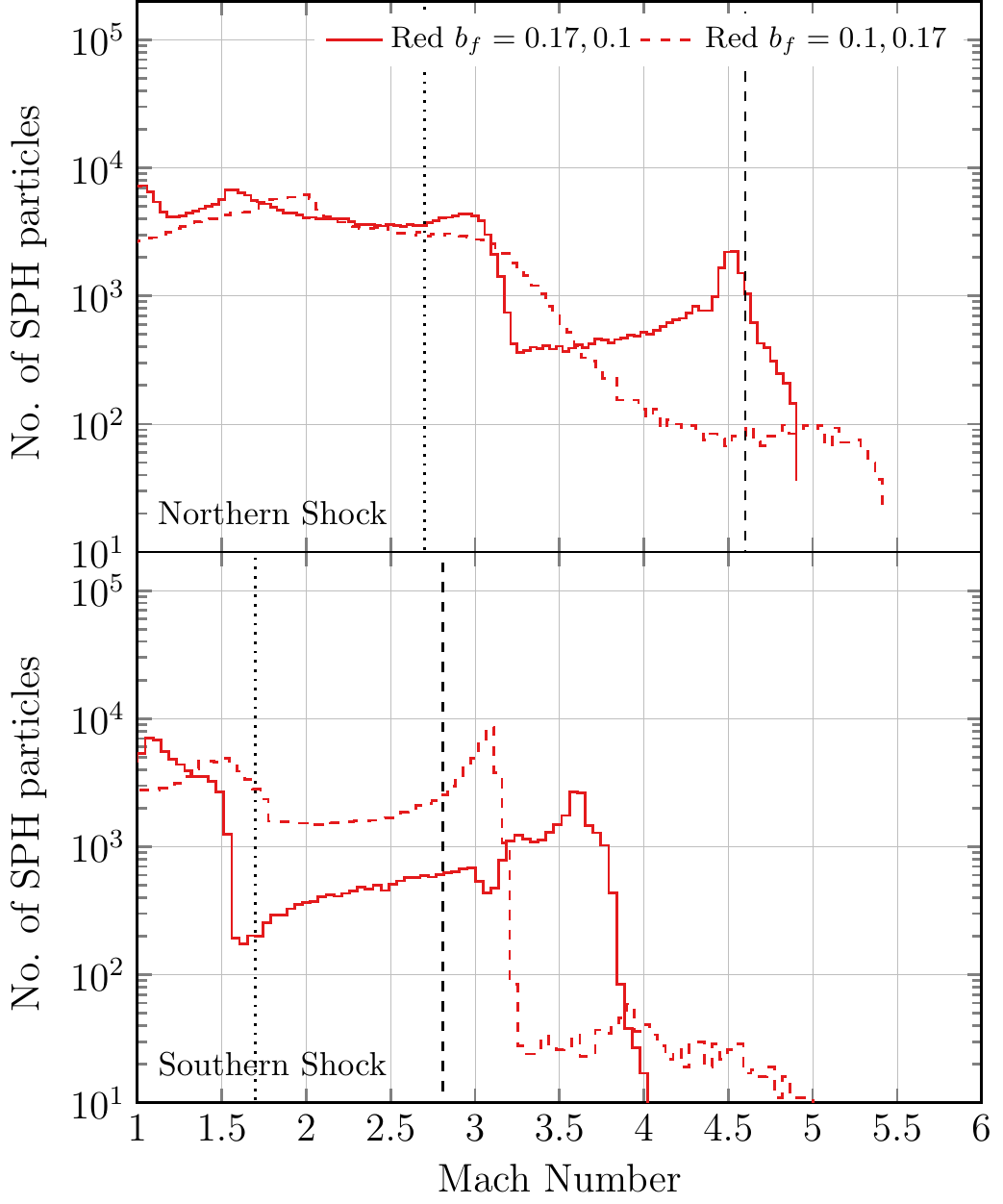}
	\caption{ Binned mach number distribution in the NS (top) and SS (bottom) from our mach finder for the Red model with $b_\mathrm{f,0} > b_\mathrm{f,1}$ (solid) and $b_\mathrm{f,0} < b_\mathrm{f,1}$ (dashed).} \label{fig.mach_bf}
\end{figure}

Here we study the influence of the relative baryon fractions of the two progenitors on the system. We  re-simulate the Red model, but reduce the baryon fraction of one progenitor from $17 \%$ to $10 \%$. The other progenitor remains unchanged. \par
We show X-ray and temperature projections of the model where the southern progenitor has reduce Baryon fraction ($b_{f,1} = 0.1$) on the left, the other one ($b_{f,0} = 0.1$) on the in figures \ref{fig.Xfast} and \ref{fig.Tfast}, panel 5 \& 6. In the first case we find that the southern shock is smaller and the northern shock larger than in the standard model. The X-ray morphology shows widening of the contact-discontinuities with respect to the standard model. The model with reduced $b_f$ in the northern progenitor shows the opposite behaviour:  The northern shock is suppressed, the southern shock enhanced and the X-ray morphology is more narrow than before. The Mach number distributions (figure \ref{fig.mach_bf}) show nearly no change in Mach number for the first case (dashed red) when compared to the standard Red model. For the second case, the suppression of the northern shock reduces the Mach number with only a few 100 particles reaching four or above. The southern shock show lower Mach numbers here, likely because the southern progenitor cannot drive the shock as efficiently anymore.\par
We conclude that the shock morphology in the observed system points toward the first case, where the southern progenitor has a Baryon fraction smaller than the northern progenitor ($b_{f,0} > b_{f,1}$). 

\subsection{Models with Substructure} \label{sect.sub}

\begin{figure}
	\centering
	\includegraphics[width=0.45\textwidth]{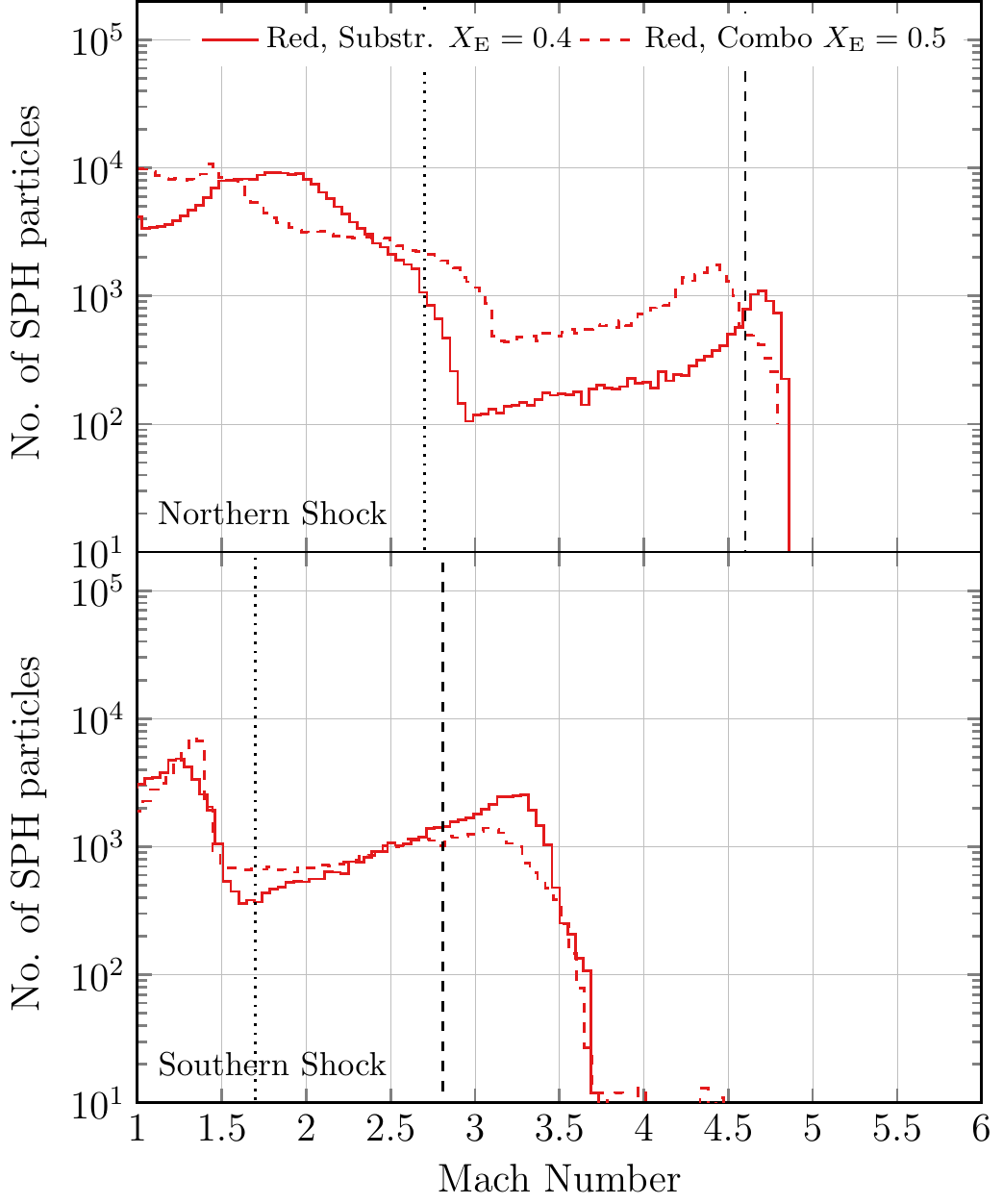}
	\caption{Binned mach number distribution in the NS (top) and SS (bottom) from our mach finder for the Red model with substructure. } \label{fig.mach_sub}
\end{figure}
\begin{figure}
	\centering
	\includegraphics[width=0.45\textwidth]{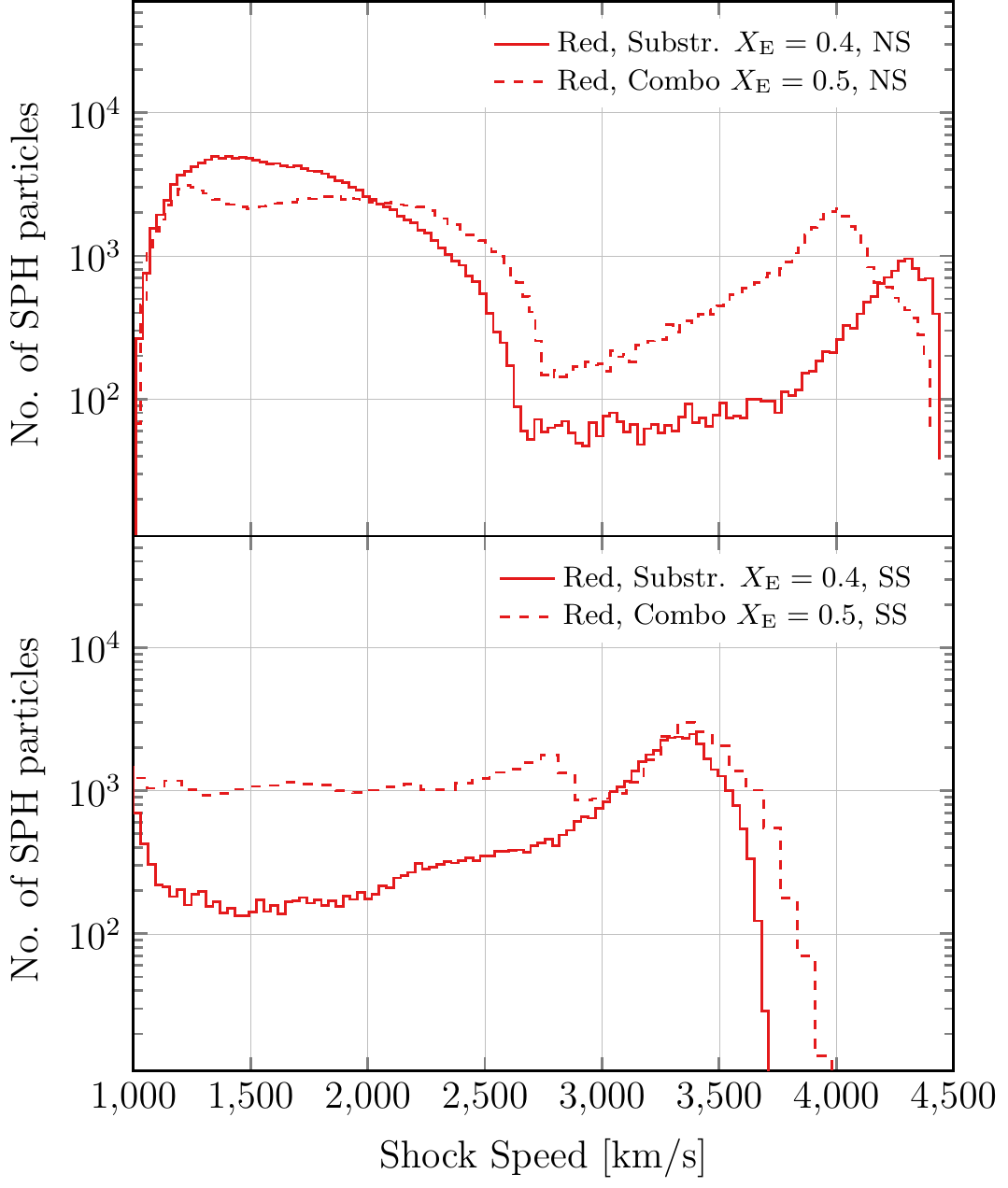}
	\caption{Binned shock velocity distribution in the NS (red, solid) and SS (red, dashed).} \label{fig.vshock_sub}
\end{figure}

We re-simulate the Red model with standard Baryon fraction $b_{f,0} = b_{f,1} = 0.17$ (red, solid) and reduced Baryon fraction in the southern progenitor $b_{f,0} = 0.17, \quad b_{f,1} = 0.10$ (red, dashed). We include a population of sub-halos in the southern progenitor in both models. Sub-halo mass distribution and spatial distribution depending on host mass are drawn from models for DM cluster substructure \citet{2010MNRAS.408..300G,2004MNRAS.352L...1G}. The halos are sampled up to the tidal radius \citep{1998MNRAS.299..728T} and are set on random orbits with the velocity limited to half the local sound speed. This allows us to investigate the influence on bulk motions and inhomogeneities in the flow on the merger state and the X-ray morphology. Of course the exact dynamics in the cluster center as well as the pre-merger state cannot claim to be a realistic model for the cluster dynamics, because they are dominated by our artificial choice of the subhalo population. In our toy model this approach is merely a proof of concept. \par 
In the beginning of the simulation, the sub-haloes of the southern progenitor fall into the cluster center,  seeding instabilities and bulk flows by gas stripping, as expected. During core passage the main DM halos interact and drive two shocks into the progenitor ICM. Shortly after the core passage most of the sub-haloes are stripped of their gas. In the first 20 Myrs after core passage, the cool core of the northern progenitor gets ablated by the additional bulk flows in the southern progenitor. Both shocks propagate into undisturbed medium, as the outer part of the souther progenitor remains undisturbed.  \par
At a shock distance of three Mpc, the observed state, both systems have $M_\mathrm{500} \approx 1.5\,\mathrm{M}_\odot$ and $R_{500} \approx 1650\,\mathrm{kpc}$, and $M_\mathrm{200} \approx 2.1\,\mathrm{M}_\odot$ and $R_{200} \approx 2500\,\mathrm{kpc}$, well in line with the weak lensing observations. \par
We show X-ray and temperature projection in figures \ref{fig.Xfast} and \ref{fig.Tfast} panel 7 \& 8. The X-rays show a elongated disturbed morphology similar to the observed cluster, where the cool core of the northern progenitor is ablated in the ICM of the southern progenitor. The DM core distance is roughly one Mpc. In the temperature map we find two well defined shocks. The NS is one Mpc (two Mpc) in size for the standard (reduced) Baryon fraction model.  Temperatures reach up to 25 keV in a narrow region behind the shock. The southern shock is larger in the standard Red model and smaller in the reduced Baryon fraction model. \par
In figure \ref{fig.mach_sub} we show Mach number histograms of the system at merger state. We find that the northern shock has a peak Mach number of 4.5, the southern shock of 3.2 in both models. We show shock speed distributions in figure \ref{fig.vshock_sub} for both models. Velocity distributions are peaked around 4100 km/s (NS) and 3300 km/s (SS). We conclude that the model with reduced Baryon fraction in the southern progenitor is most consistent with the radio scenario and the majority of observations.

\subsection{Models for the X-ray Scenario}

\begin{figure*}
	\centering
	\includegraphics[width=0.45\textwidth]{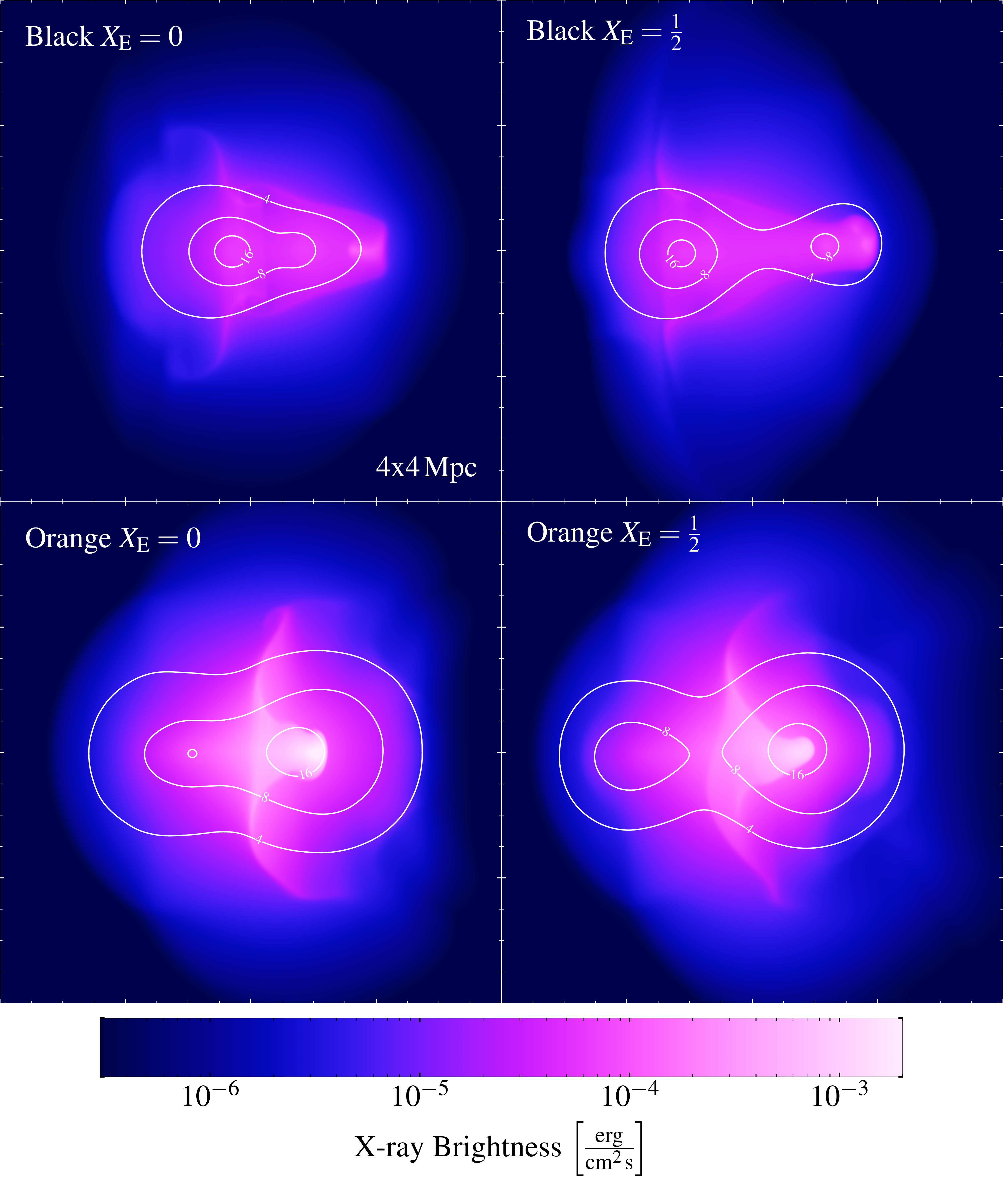}
	\includegraphics[width=0.45\textwidth]{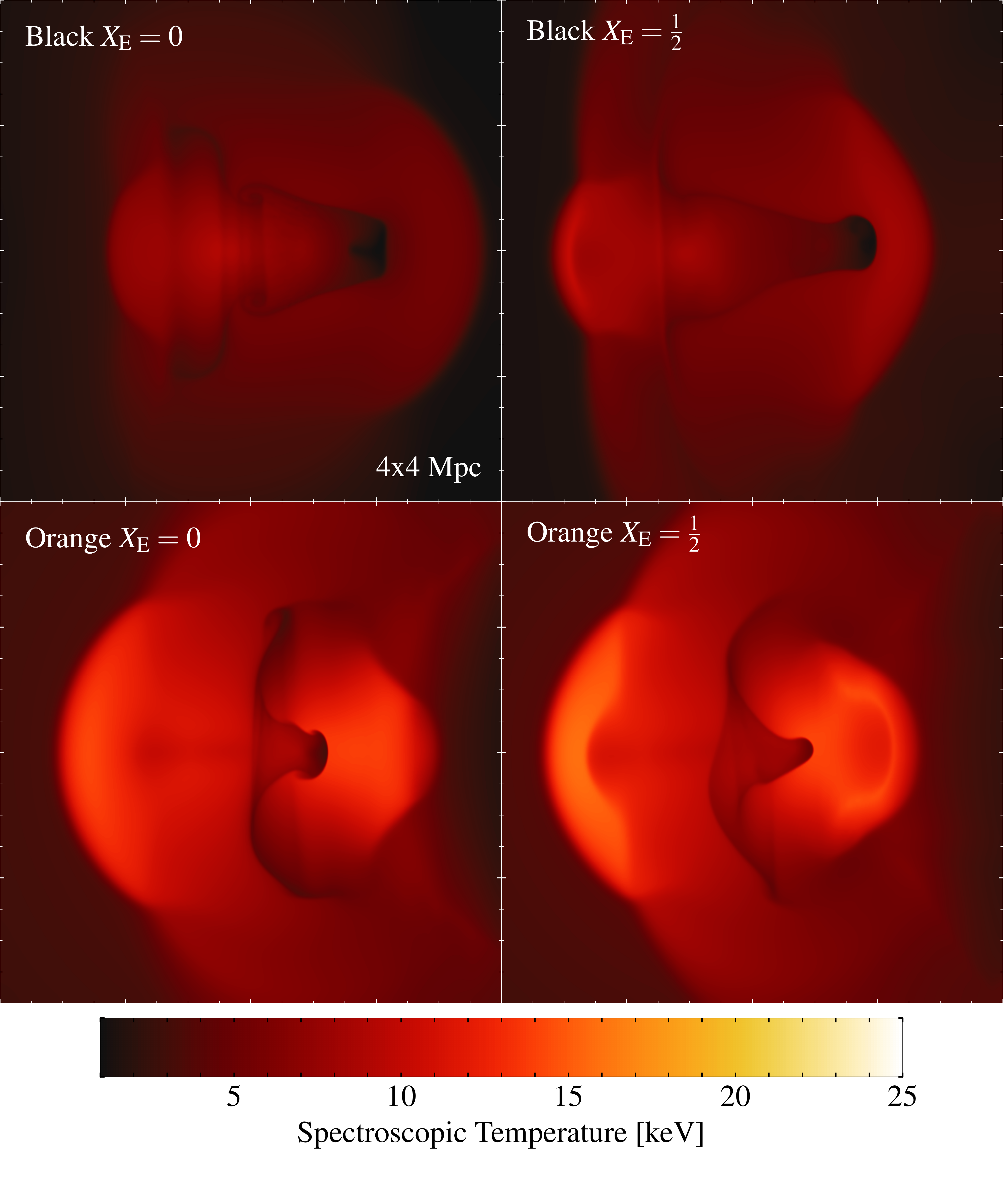}
	\caption{Left: Projected X-ray emissivity in erg/cm$^2$/Hz/s of models Black and Orange, with and without initial velocity. We overplot contours of the DM mass distribution in $10^{21} \,\mathrm{g}/\mathrm{cm}^2$. Right: Spectroscopic temperature of the same models.} \label{fig.TXextra}
\end{figure*}
\begin{figure}
	\centering
	\includegraphics[width=0.45\textwidth]{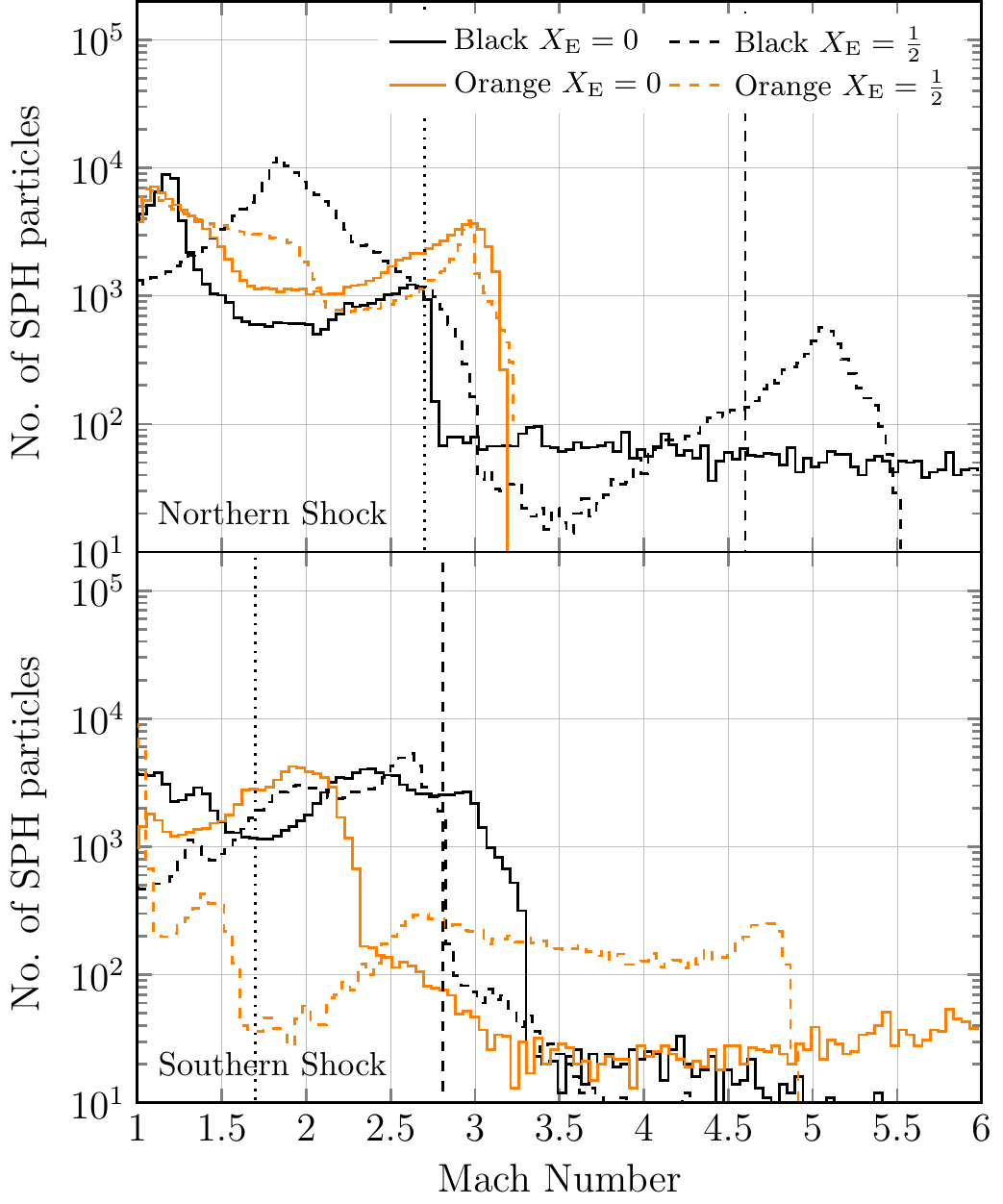}
	\caption{ Binned mach number distribution in the NS (top) and SS (bottom) from our mach finder for the Orange and Black model with $X_\mathrm{E} = 0$ (solid) and $X_\mathrm{E} = 0.5$ (dashed).} \label{fig.mach_extra}
\end{figure}

In this section we explore two model classes further outside the observed parameter range to force slower speeds in the NS : A model with a mass ratio smaller than one (Orange) and a model with large cut-off radius (Black). Both lead to a reduced Mach number in the NS. The Black mode allows a lower total mass and thus lower potential energy in the system . The Orange model decreases the mass of the progenitor driving the NS, at the cost of obtaining a higher Mach number in the SS. \par
We again show X-ray and temperature projections in figure \ref{fig.TXextra} and Mach number distribution in figure \ref{fig.mach_extra}. Basic model parameters can be found in tables \ref{tab.models} and \ref{tab.models_fast}. \par

\subsubsection{A Model with $R_\mathrm{cut} \gg r_{200}$}

To test the limits of what is allowed with our cluster model, we simulate a model with a very large value for $R_\mathrm{cut} \rightarrow \infty$  (Black model, figure \ref{fig.TXextra}, top row). The Black model with large cut-off radius shows structural properties similar to the Red model, however at a lower X-ray luminosity ($L_\mathrm{X} \approx 2 \times 10^{44} \,\mathrm{erg} /\mathrm{cm}^2/ \mathrm{s}/\mathrm{Hz}$). We find a triangular shape in the X-rays with a core distance of 500 kpc in the slow and 1 Mpc in the fast variant. ICM temperatures range from 10 to 20 keV, with the NS again standing out with 25 keV and the SS with 20 keV. Even though shock velocities are lower compared to the Red model, 3000 (NS) and  Mach numbers in the faster variant of 5 (NS) and 2.7 (SS) are surprisingly consistent with the radio scenario, not the X-ray scenario. Shock speeds are slower than in the radio scenario though, 3200 km/s (NS) and 2040 km/s (SS), as expected from a low mass model. We conclude that the Black model is not a good fit to either of the scenarios. In our model, lower mass systems cannot be consistent with the X-ray temperature upstream of the shocks.

\subsubsection{A Model with $X_\mathrm{M} < 1$}

Finally, we simulate a model with inverted mass ratio $X_\mathrm{M} = \frac{1}{2}$ (Orange). We find a concentrated X-ray morphology with two characteristic contact discontinuities in the center of the system (figure \ref{fig.TXextra}, bottom four panels). DM peak separation in the slow and the fast variant for the model is one Mpc. Temperatures in the center of the cluster are roughly 10 keV, but reach again 25 keV in both shocks. The NS is larger than 2 Mpc in both variants, while the SS has an elongated size of only one Mpc. Mach numbers in the NS are roughly three in both variants. However, in the SS we find a Mach number of 2.2 in the slow model and 4.6 in the fast model. This is confirmed by the mach number distributions (figure \ref{fig.mach_extra}. We note that the SS in the slow model is not fully developed and shows a long tail of particles with high Mach numbers of up to six. A model with inverted mass ratio is consistent with some aspects of the X-ray scenario (Mach number in the NS, shock speeds in both shocks), but is inconsistent with other properties (shock temperature, shock morphology, SS properties). Hence, we conclude that none of our models are an acceptable fit to the X-ray scenario of the two shocks.

\section{Discussion} \label{sect.discussion}

\subsection{The Radio Scenario}

Using our simulations we have found a class of models that is widely consistent with the radio scenario. Assuming a cut-off radius consistent with observations of the Perseus cluster, simulations with a combined progenitor mass of $M_\mathrm{tot} = 1.5 - 2\times 10^{15} M_\odot$, and a mass ratio of $X_\mathrm{M} = 1.5 - 2.5$ generate shocks consistent with the observed radio relics (models Red, Yellow, Brown, Pink). Lower combined progenitor masses require larger cut-off radii ($r_\mathrm{cut}$) to match the upstream shock properties, with lowest masses of around $0.75 \times 10^{15} \,\mathrm{M}_\odot$ (model Black).   \par
The system is further constrained by the X-ray brightness, which suggests a combined progenitor mass around $M_\mathrm{tot} = 1.5 \times 10^{15} M_\odot$, mostly dependent on the slope of the beta profiles (compare models Red and Pink) and the assumed baryon fraction. As noticed before (Hoang et al. submitted to MNRAS), the system is under-luminous, which is consistent with our best-fit model that has very flat ICM density profiles and where the southern progenitor has a low Baryon fraction in the ICM (Hoang et al. submitted to MNRAS). \par
In simulations consistent with the radio scenario, the shock that forms the northern relic travels outward along the merger axis. It is collimated by the contact discontinuities formed between the  merging two ICM's. In the fiducial parameter region, shock speeds in the simulations lie around 4000 km/s, Mach numbers between four and five. This is true over a wide range of masses, compare Mach numbers from Black \& Red with $X_\mathrm{E} = 0.5$. This is because lower merger masses result in lower shock speeds, but also lower upstream sound speeds. Thus the Mach number in the shocks is only weakly dependent on mass. \par
Temperatures in the NS reach 20-25 keV in a $< 200$ kpc region behind the shock, as expected from an upstream temperature of 3 keV and a Mach number above four. We note that our simulations could well be resolution limited here, i.e. the true high temperature region is likely even smaller. We defer a resolution study to future work. The simulated southern shock shows a lower Mach number of 3-3.5, with speeds of about 3200 km/s and again temperatures of 15-20 keV, also roughly consistent with the radio scenario. Our approach is not able to model the complex bulk flows, shock structure and galaxy shock interaction taking place in the SR. In particular, emission from radio galaxies in the SR can steepen the spectral index of the relic and decrease the Mach number inferred from the radio spectrum of the SR. We note that the large size of the SS in the simulations (2 Mpc) is indeed roughly consistent with new low resolution LOFAR data of the SR (Hoang et al. submitted to MNRAS).  \par
The main discrepancy of our simulations with the observed radio scenario is the exact timing of the merger state, i.e. the NS tends to be too small when both shocks have a distance of three Mpc. We were able to show that different relative Baryon fraction could account for this (compare the two substructure models). The small $\beta$ parameter and the X-ray brightness also suggest that the southern progenitor likely has a lower Baryon fraction than simulated, so the mass of the southern halo is likely larger than in our fiducial model.   We set to the canonical cosmological value inside $r_{200}$ of $b_\mathrm{f} = 17\% \approx \frac{\Omega_\mathrm{b}}{\Omega_\mathrm{DM}}$ \citep[e.g.][]{2006ApJ...640..691V,2013ApJ...778...14G,2014A&A...571A..16P}. However, this parameter varies significantly among observed clusters, depending on how much of the ICM has been converted into stars in galaxies since the last major merger (Perseus: $b_\mathrm{f} = 23\%$ \citep{2011Sci...331.1576S}, Cygnus: $b_\mathrm{f} \le 10 \%$ (Halbesma et al. in prep.)). \par
Given the number of parameters and their error bars we do not attempt to refine our models even further. We have likely reached the limit of what a toymodel can achieve in reproducing a complex merging cluster. 

\subsection{The X-ray Scenario}

We were not able to model the X-ray scenario using our simulations in a satisfactory way. Nonetheless, low mass models generally reproduce some aspects of the X-ray scenario. The Black model without initial velocity shows a Mach number of 2.6 and a shock velocity around 2000 km/s in the NS. However, the SS is too fast with Mach numbers of about three and velocities of 2000 km/s, when 1.7 (!) is inferred from the X-ray observations. Giving the merger an initial velocity results in a NS consistent with the radio scenario, the model is ''unstable'' under this assumption. \par
The Orange model without initial velocity does a little better, with Mach numbers of about three in the NS and about two in the SS. Again, these values change drastically once the clusters are set on a non-zero orbit. Furthermore, the small SS in the simulation is inconsistent with recent observations of the cluster with LOFAR  (Hoang et al. submitted to MNRAS). \par
Leaving this aside, it seems that we would need to reduce the total mass of the system even further, to allow for an initial velocity to fit the X-ray scenario (less than model Blue). However, Black and Orange models are already very under-luminous in the X-rays. Steeper $\beta$ profiles for the ICM would increase the X-ray luminosity of a low mass system, but decrease the upstream temperatures and thus increase the Mach number. This makes it not possible for us to fit X-ray luminosity and upstream temperature simultaneously. This circular argument suggests that the X-ray scenario becomes inconsistent with itself at low masses. \par
All simulations show temperatures  too high to be consistent with the X-ray scenario behind the shocks. In shocks with consistent Mach numbers (Orange and Black) this points to an excess in upstream temperature or shock velocity. Our simulations assume the ICM behaves like a single fluid in the shocked region, which might very well be incorrect. One solution for the Mach number inconsistency could be a two temperature structure of the ICM behind the shock. Here the thermal electrons, which are visible in the X-rays would have a temperature different from the thermal ions that mediate the shock. This has been observed on galaxy scales in the centre of clusters \citep{2012ApJ...749..186G}, but not been detected in shocks in the outskirts \citep{2012MNRAS.423..236R,2006ESASP.604..723M}. \par
We note that the fluid assumption for the ICM, as derived from Coulomb collisions in the Spitzer model \citep{1956pfig.book.....S,1988xrec.book.....S} is indeed not valid in the region of the Sausage relic. Given the physical shock properties assumed here, Coulomb mean free paths of thermal ions in the NS are about $d_\mathrm{mfp,C} \approx 200\,\mathrm{kpc}$ ahead of the relic. We would then expect a shock thickness of at least of that order. However, the \emph{relic} thickness at 610 MHz is observed around 50 kpc. As the relic emission is caused by cooling cosmic-ray electrons, the injection region has to be at least an order of magnitude smaller than its extend to result in the observed steepening of the spectral index profile. Thus another process has to maintain collisionality of thermal protons and cosmic-rays in the shock and ahead of the relic. It is unclear why the thermal electrons should not be heated, but cosmic-ray electrons seem to be efficiently injected or re-accelerated. \par
If on the other hand collionality in the ICM is maintained through the interaction of particles with plasma waves, the fluid approximations is valid far below kpc scale \citep[e.g. ][ and references therein]{2014IJMPD..2330007B,2014MNRAS.443.3564D}. In this case, we would expect high temperatures for the thermal electrons behind the shock as well and other effects are responsible for the inconsistency in Mach number. If indeed radio halos are caused by turbulent reacceleration \citep{2001ApJ...557..560P}, they are direct evidence for this mechanism acting on radio dark cosmic-ray electrons.

\subsection{Merger Speed vs. Shock Speed}

\begin{figure}
	\centering
	\includegraphics[width=0.45\textwidth]{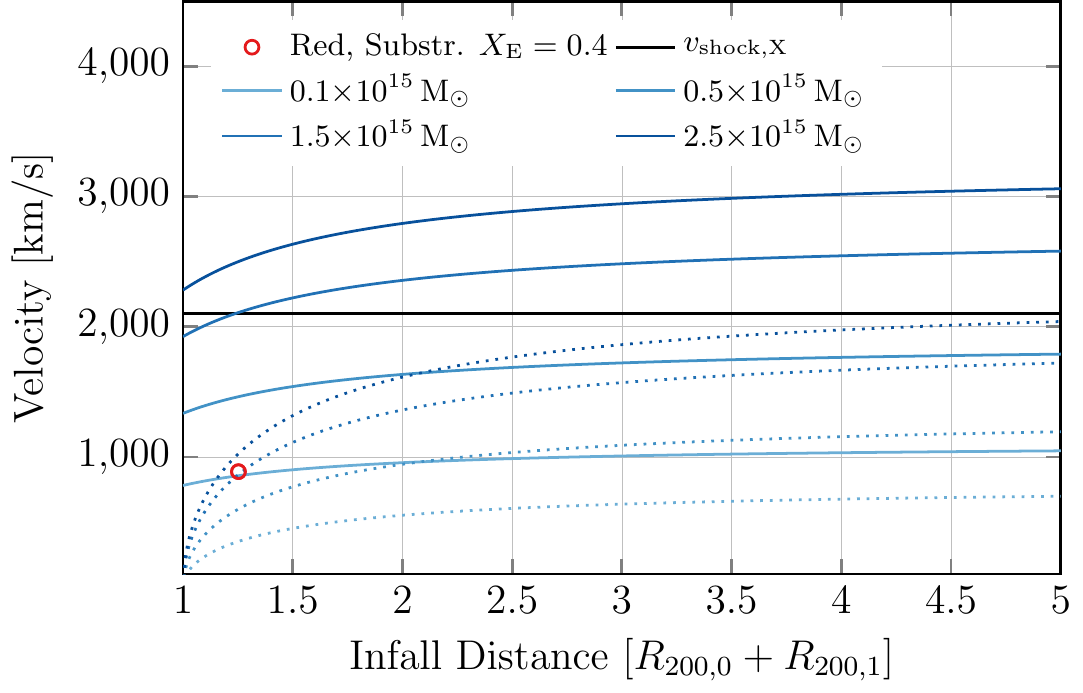}
	\caption{Combined infall velocity  at the distance of the two virial radii in km/s over merger distance in units of that distance.} \label{fig.kinematics}
\end{figure}

Shock speeds in the radio scenario are only reproduced for orbits with a zero energy fraction of $X_\mathrm{E} \ge 0.4$ (compare Red with and without initial velocities).  This is not surprising given that clusters likely detach from the Hubble flow before their virial radii touch. Following \citet{2002ASSL..272....1S}, the combined in-fall velocity $v_0$ of the progenitors at a distance $d$ with impact parameter $b_0$ before the merger, is connected to the distance $d_0$ at which both progenitors began to merge:
\begin{align}
	v_0 &\approx 2930 \sqrt{\frac{M_0 + M_1}{10^{15}\,\mathrm{M}_\odot}} \left(\frac{d}{1\,\mathrm{Mpc}}\right)^{-\frac{1}{2}} \left[ \frac{1 - \frac{d}{d_0}}{1 - \frac{b^2}{d_0^2}} \right]^{\frac{1}{2}} \,\frac{\mathrm{km}}{\mathrm{s}}. \label{eq.v0}
\end{align}
For the fiducial model with substructure (Red), we then find an initial separation of 4500 Mpc, $30\%$ larger than the combined virial radius of the progenitors. \par
We can compare the speeds of the merger with the shock speeds in the two scenarios: We plot the in-fall velocity $v_0$ at a distance of combined virial radii (dotted) and half that value (solid, about two Mpc) in figure \ref{fig.kinematics} for a mass range of $0.1-2.5\times10^{15}\,\mathrm{M}_\odot$. We add the fiducial model as red circle, the velocity is independent of mass ratio. We over-plot the shock velocity in the X-ray scenario as black dashed line. \par
In-fall velocities are only a fraction of the shock speed  in the radio scenario (4200 km/s) for all relevant mass ranges, even at a distance of half the combined virial radii. This is consistent with the simulations, where a large fraction of the shock kinetic energy is generated during core passage, when the DM cores rapidly interact and drive two shocks outwards. \par
In contrast, the shock speed of the X-ray scenario, 2100 km/s, is exceeded at cluster in-fall for all combined masses larger than $10^{15} \,\mathrm{M}_\odot$ at a distance of half the virial radius (two to three Mpc). Any additional kinetic energy by e.g. DM core interaction will lead to shock velocities too high to be consistent with the X-ray scenario. This kinetic argument motivates why we would need very small cluster masses ($M_0 + M_1 \approx 10^{14} \, M_\odot$) to reproduce the shock speeds in the X-ray scenario. Cluster infall speeds alone are inconsistent with the shock speeds estimated in the X-ray scenario for systems consistent with weak lensing observations. \par

\subsection{Cluster Masses vers. Merger State} \label{sect.scaling}

In our approach, clusters scale with combined progenitor mass, so morphological properties remain widely identical over  a mass range, holding $X_\mathrm{E}$, $b_f$, $\beta$ and the mass ratio constant (compare Red and Black). However, the dynamical state of the system at observation time changes with cluster mass, because the characteristic velocities (eq. \ref{eq.v0}) and length scales of the system ($r_{200}$) change. However, the observed shock distance remains the same (3 Mpc). The most notable effect is that the distance of the DM mass peaks decreases with decreasing cluster mass (keeping $X_\mathrm{E}$ constant). The second core passage is occuring at the observed state for masses of about $10^{14} \,M_\odot$ (if the system is started at rest, see model Black with $X_E = 0$). \par
In the slow low mass simulations this results in a core distance significantly smaller than observed (Black, Purple). In these models the DM cores have turned around and are shortly before second core passage, when the shocks are at 3 Mpc distance. This sets a lower limit on the total mass of the system, which is roughly $10^{15} \,M_\odot$ for $X_\mathrm{E} = 0$. The slowest allowed model (Green) still predicts shock velocities  inconsistent with the X-ray scenario. This suggests that the X-ray scenario is not consistent with the merger state of the observed system, as inferred from the weak lensing mass peaks.

\subsection{Time Scales vers. Radio Halo}

With decreasing system masses and merger velocities, merger time scales increase. We find that roughly 500-750 Myr lie between first core passage and observed state in the fiducial mass range (Red, Green). This is consistent with the existence of the flat-spectrum giant radio halo found in the system (Hoang et al submitted to MNRAS). Giant radio haloes take roughly 500 Myrs to ''switch on'' \citep{2013MNRAS.429.3564D}. In the slow, low mass models and the X-ray scenario for the NS (Purple, Black), the travel times are larger, about 750-1000 Myrs. These shock travel times become inconsistent with a flat-spectrum radio halo. Because the cosmic-ray electrons had time to cool, we would expect a steep spectrum radio halo in this case. Thus time scale arguments exclude low mass models as good matches to the observations. \par

\subsection{Guiding Future Observations}

\begin{figure}
	\centering
	\includegraphics[width=0.45\textwidth]{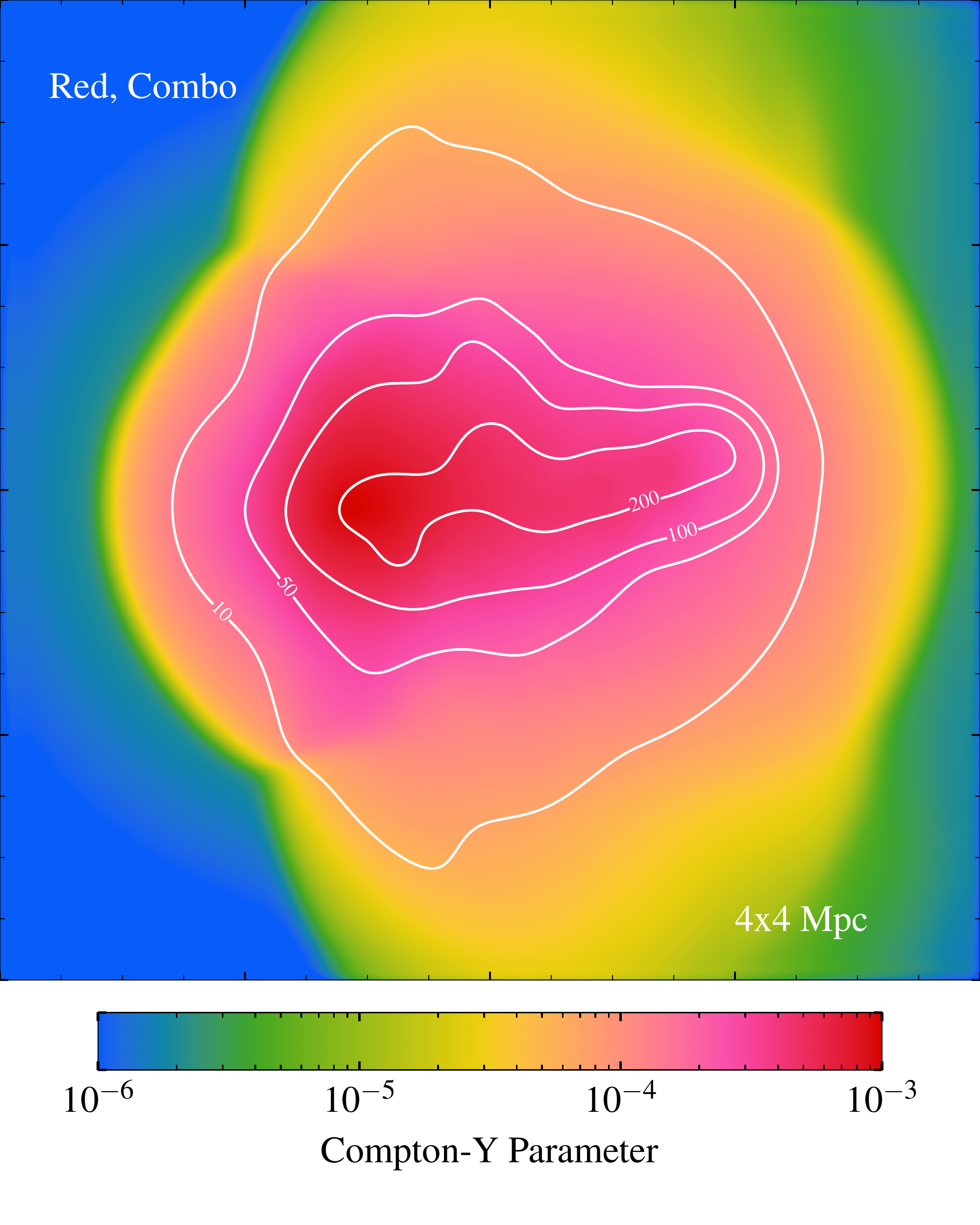}
	\caption{Projected Compton-Y parameter from the Red model with substructure and reduced Baryon fraction. In contours X-ray brightness in $10^{-6}$ erg/s/cm$^2$ in the ROSAT band.} \label{fig.proj_CY}
\end{figure}

Our results strongly suggest to use X-ray observatories to search for excess temperatures in the downstream region of the northern shock of  CIZA J2242.8+5301 and shocks in other cluster with a radio derived Mach number larger than 4. Our numerical models, combined with current observations, including the total X-ray brightness from ROSAT, clearly favour the radio scenario, with high masses, high Mach numbers and high shock velocities. Among the simulations, post-shock temperatures regularly exceed 15 keV, however in a small area on the sky with very low X-ray emissivity. Thus a long exposure with the optimal instrument might be necessary to find a high temperature component. We plan to use our fiducial model with simulators for X-ray observatories (Marx, SIXTE etc.) to find the optimal instrument and exposure parameters for this search in a future paper. \par
Another clue to the total mass of the system could come from the SZ-effect of the total cluster. The SZ-mass scaling relation is reasonably tight to confirm the weak lensing mass measurements. We are currently not aware of a successful measurement of the Compton-Y parameter in CIZA J2242.8+5301 \citep{2014MNRAS.441L..41S,2016A&A...591A.142B}. We show a prediction of the projected Compton-Y parameter from the fiducial model in figure \ref{fig.proj_CY}. We over-plot the X-ray emission in white contours in units of $10^{-6}$ erg/s/cm$^2$

\section{Conclusions}\label{sect.conclusions}

We attempted to find a model for the merging galaxy cluster CIZA J2242.8+5301 and its two prominent shocks. We showed analytically that in the northern shock, upstream X-ray temperatures and radio properties are consistent with each other. These values are in turn consistent with weak lensing cluster masses, assuming a reasonable model for the underlying merger progenitors. We then explored the resulting parameter space using idealized simulations of galaxy cluster mergers, where we modeled the northern progenitor as a cool core. We found that models with a combined progenitor mass of $M_\mathrm{tot,200} = 1.5-2\times10^{15}\,\mathrm{M}_\odot$ and mass ratios between 1.5 and 2.5 are consistent with the majority of observations. In particular:
\begin{enumerate}
	\item X-ray brightness, morphology and upstream shock temperature.
	\item Weak lensing total mass, sub-cluster mass ratio and location of the mass peaks.
	\item Size of the shocks, their radio inferred Mach numbers and shock speeds.
\end{enumerate}
We were not able to find a model that is consistent with all the X-ray observations, in particular the observed downstream temperatures. Even the most conservative models for the cluster dynamics show shock speeds and thus downstream temperatures larger than observed. In particular :
\begin{enumerate}
	\item Kinetic arguments suggest that X-ray derived shock speeds cannot be reproduced for models with masses above $0.5\times10^{15}\,M_\odot$.
	\item Models with masses below $0.5\times10^{15}\,M_\odot$ are under-luminous in the X-rays.
	\item The observed DM core distance is inconsistent with simulations that reproduce X-ray derived shock speeds.
	\item Time scale arguments suggest that the observed flat spectrum radio halo is inconsistent with the X-ray derived shock speeds. 
\end{enumerate}
We concluded that the radio scenario is preferred by the observations and simulations. An extensive search for a high temperature component in the northern shock using X-ray observatories should be able to rule out the X-ray scenario. \par

\section{Acknowledgements}\label{sect.ack}

The authors thank Aurora Simionescu, Irina Zhuravleva and Ondrej Urban for providing the observed profiles of the Perseus cluster. JD thanks Carlo Giocoli for extensive discussions about DM substructure in clusters. We thank Tom Jones for discussions and access to the Itasca cluster at MSI for this research. The research leading to these results has received funding from the People Programme (Marie Sklodowska Curie Actions) of the European Union’s Eighth Framework Programme H2020 under REA grant agreement no 658912, ''Cosmo Plasmas''. HR acknowledges support from the ERC Advanced Investigator programme NewClusters 321271. We used Julia\footnote{www.julialang.org} and PGFPlots for all graphics in this document. All colormaps are perceptually uniform (\citet{2015arXiv150903700K} , colorbrewer.org).

\appendix

\section{A Non-Cool-Core Model} \label{app.nonCC}

\begin{figure*}
	\centering
	\includegraphics[width=0.45\textwidth]{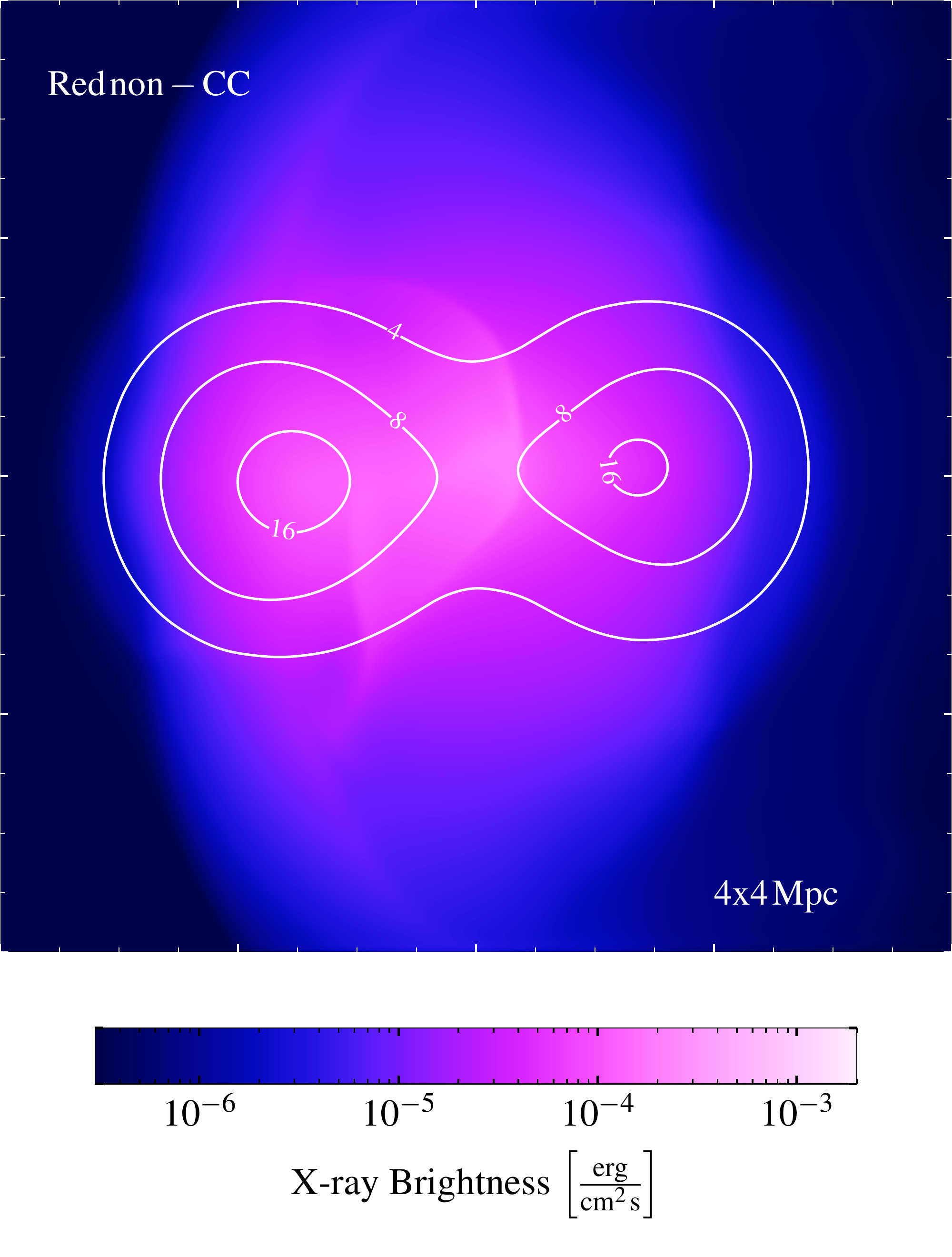}
	\includegraphics[width=0.45\textwidth]{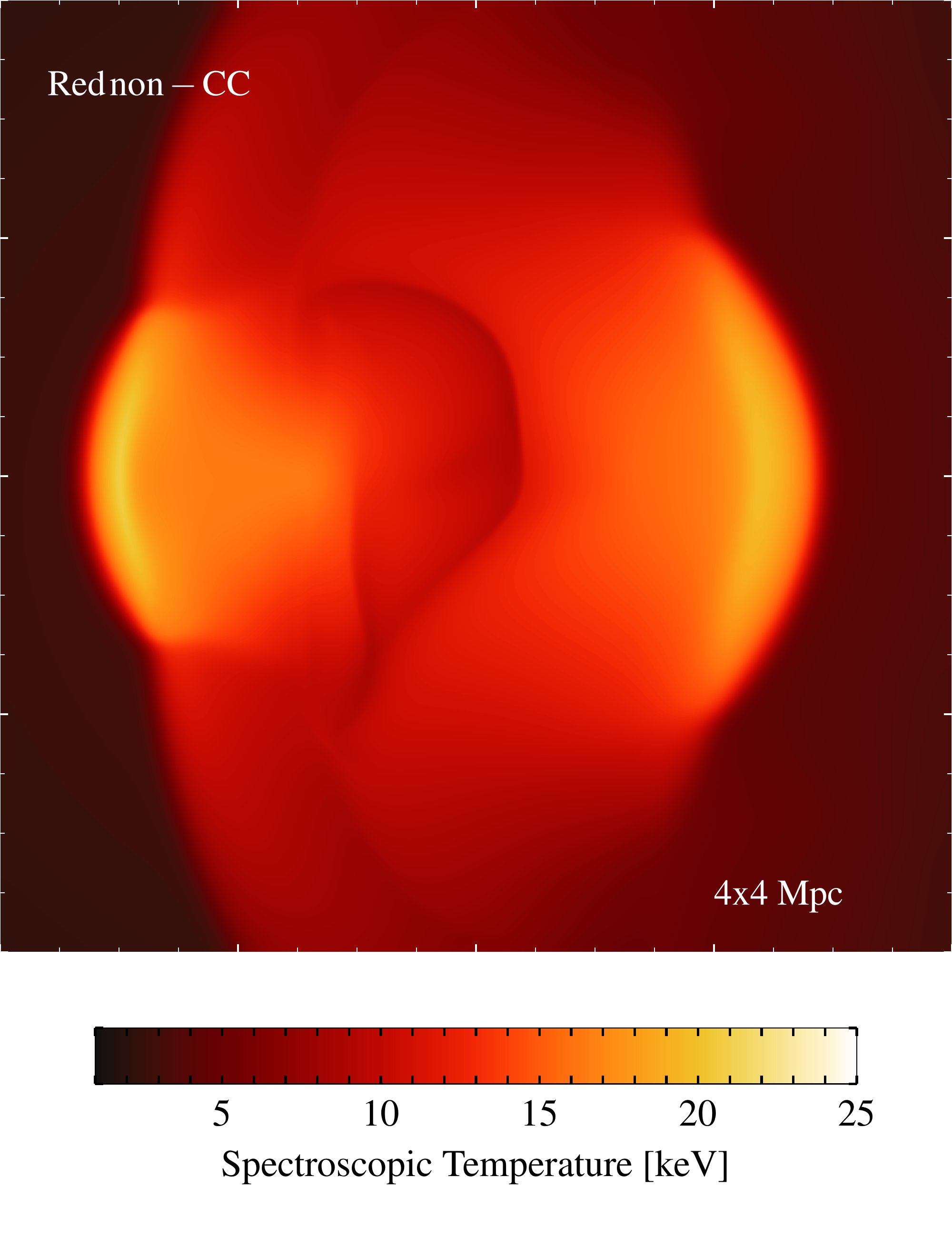}
	\caption{Left: Projected X-ray emissivity in erg/cm$^2$/Hz/s of models Red without initial velocity, but without a cool core. We overplot contours of the DM mass distribution in $10^{21} \,\mathrm{g}/\mathrm{cm}^2$. Right: Spectroscopic temperature of the same models.} \label{fig.proj_nonCC}
\end{figure*}

We present a model with the same parameter configuration as the Red model (section \ref{sect.fiducial}), but as a non-cool-core merger. I.e. the northern progenitor has $r_\mathrm{core} = r_\mathrm{s}/3$. In figure \ref{fig.proj_nonCC} we show X-ray (left) and temperature projections (right) of the system at the observed state. The morphology of the system in the X-ray emission is less elongated and more diffuse than in the standard Red model, with a characteristic contact discontinuity in the centre of the cluster. The temperature structure in the shocks is similar to the standard Red model, but the characteristic elongated structure of the observed cluster is not present. X-ray luminosity, Mach numbers and dynamical parameters are roughly the same as in the standard Red model. We conclude that a non-CC merger is not favoured by the observed X-ray morphology of the cluster.

\bibliographystyle{mn2e} \bibliography{master}

\begin{thebibliography}{}

\bibitem[\protect\citeauthoryear{{Akamatsu} \& {Kawahara}}{{Akamatsu} \&
  {Kawahara}}{2013}]{2013PASJ...65...16A}
{Akamatsu} H.,  {Kawahara} H.,  2013, \pasj, 65, 16

\bibitem[\protect\citeauthoryear{{Akamatsu}, {van Weeren}, {Ogrean},
  {Kawahara}, {Stroe}, {Sobral}, {Hoeft}, {R{\"o}ttgering}, {Br{\"u}ggen} \&
  {Kaastra}}{{Akamatsu} et~al.}{2015}]{2015AA...582A..87A}
{Akamatsu} H.,  {van Weeren} R.~J.,  {Ogrean} G.~A.,  {Kawahara} H.,  {Stroe}
  A.,  {Sobral} D.,  {Hoeft} M.,  {R{\"o}ttgering} H.,  {Br{\"u}ggen} M.,
  {Kaastra} J.~S.,  2015, \aap, 582, A87

\bibitem[\protect\citeauthoryear{{Barnes}}{{Barnes}}{2012}]{2012MNRAS.425.1104B}
{Barnes} J.~E.,  2012, \mnras, 425, 1104

\bibitem[\protect\citeauthoryear{{Basu}, {Vazza}, {Erler} \& {Sommer}}{{Basu}
  et~al.}{2016}]{2016A&A...591A.142B}
{Basu} K.,  {Vazza} F.,  {Erler} J.,    {Sommer} M.,  2016, \aap, 591, A142

\bibitem[\protect\citeauthoryear{{Beck}, {Dolag} \& {Donnert}}{{Beck}
  et~al.}{2016}]{2016MNRAS.458.2080B}
{Beck} A.,  {Dolag} K.,    {Donnert} J.,  2016, \mnras, 458, 2080

\bibitem[\protect\citeauthoryear{{Beck}, {Murante}, {Arth}, {Remus}, {Teklu},
  {Donnert} \& {Planelles}}{{Beck} et~al.}{2016}]{2016MNRAS.455.2110B}
{Beck} A.,  {Murante} G.,  {Arth} A.,  {Remus} R.,  {Teklu} A.,  {Donnert} J.,
    {Planelles} S. e.~a.,  2016, \mnras, 455, 2110

\bibitem[\protect\citeauthoryear{{Binney} \& {Tremaine}}{{Binney} \&
  {Tremaine}}{2008}]{2008gady.book.....B}
{Binney} J.,  {Tremaine} S.,  2008, {Galactic Dynamics: Second Edition}.
Princeton University Press

\bibitem[\protect\citeauthoryear{{Br{\"u}ggen}, {Bykov}, {Ryu} \&
  {R{\"o}ttgering}}{{Br{\"u}ggen} et~al.}{2012}]{2012SSRv..166..187B}
{Br{\"u}ggen} M.,  {Bykov} A.,  {Ryu} D.,    {R{\"o}ttgering} H.,  2012, \ssr,
  166, 187

\bibitem[\protect\citeauthoryear{{Brunetti} \& {Jones}}{{Brunetti} \&
  {Jones}}{2014}]{2014IJMPD..2330007B}
{Brunetti} G.,  {Jones} T.~W.,  2014, International Journal of Modern Physics
  D, 23, 30007

\bibitem[\protect\citeauthoryear{{Burns}, {Loken}, {Roettiger}, {Norman} \&
  {Clarke}}{{Burns} et~al.}{1993}]{1993LNP...421..267B}
{Burns} J.~O.,  {Loken} C.,  {Roettiger} K.,  {Norman} M.~L.,    {Clarke}
  D.~A.,  1993, in {R{\"o}ser} H.-J.,  {Meisenheimer} K.,  eds, Jets in
  Extragalactic Radio Sources Vol.~421 of Lecture Notes in Physics, Berlin
  Springer Verlag, {Jet Disruptions at the Cores of Rich Galaxy Clusters}.
p.~267

\bibitem[\protect\citeauthoryear{{Cavaliere} \& {Fusco-Femiano}}{{Cavaliere} \&
  {Fusco-Femiano}}{1976}]{1976A&A....49..137C}
{Cavaliere} A.,  {Fusco-Femiano} R.,  1976, \aap, 49, 137

\bibitem[\protect\citeauthoryear{{Dawson}, {Jee}, {Stroe}, {Ng}, {Golovich},
  {Wittman}, {Sobral}, {Br{\"u}ggen}, {R{\"o}ttgering} \& {van
  Weeren}}{{Dawson} et~al.}{2015}]{2015ApJ...805..143D}
{Dawson} W.~A.,  {Jee} M.~J.,  {Stroe} A.,  {Ng} Y.~K.,  {Golovich} N.,
  {Wittman} D.,  {Sobral} D.,  {Br{\"u}ggen} M.,  {R{\"o}ttgering} H.~J.~A.,
  {van Weeren} R.~J.,  2015, \apj, 805, 143

\bibitem[\protect\citeauthoryear{{Diehl}, {Rockefeller}, {Fryer}, {Riethmiller}
  \& {Statler}}{{Diehl} et~al.}{2012}]{2012arXiv1211.0525D}
{Diehl} S.,  {Rockefeller} G.,  {Fryer} C.~L.,  {Riethmiller} D.,    {Statler}
  T.~S.,  2012, ArXiv e-prints

\bibitem[\protect\citeauthoryear{{Dolag} \& {Stasyszyn}}{{Dolag} \&
  {Stasyszyn}}{2009}]{2009MNRAS.398.1678D}
{Dolag} K.,  {Stasyszyn} F.,  2009, \mnras, 398, 1678

\bibitem[\protect\citeauthoryear{{Donnert} \& {Brunetti}}{{Donnert} \&
  {Brunetti}}{2014}]{2014MNRAS.443.3564D}
{Donnert} J.,  {Brunetti} G.,  2014, \mnras, 443, 3564

\bibitem[\protect\citeauthoryear{{Donnert}, {Dolag}, {Brunetti} \&
  {Cassano}}{{Donnert} et~al.}{2013}]{2013MNRAS.429.3564D}
{Donnert} J.,  {Dolag} K.,  {Brunetti} G.,    {Cassano} R.,  2013, \mnras, 429,
  3564

\bibitem[\protect\citeauthoryear{{Donnert}}{{Donnert}}{2014}]{2014MNRAS.438.1971D}
{Donnert} J.~M.~F.,  2014, \mnras, 438, 1971

\bibitem[\protect\citeauthoryear{{Donnert}, {Stroe}, {Brunetti}, {Hoang} \&
  {Roettgering}}{{Donnert} et~al.}{2016}]{2016MNRAS.462.2014D}
{Donnert} J.~M.~F.,  {Stroe} A.,  {Brunetti} G.,  {Hoang} D.,    {Roettgering}
  H.,  2016, \mnras, 462, 2014

\bibitem[\protect\citeauthoryear{{Duffy}, {Schaye}, {Kay} \& {Dalla
  Vecchia}}{{Duffy} et~al.}{2008}]{2008MNRAS.390L..64D}
{Duffy} A.~R.,  {Schaye} J.,  {Kay} S.~T.,    {Dalla Vecchia} C.,  2008,
  \mnras, 390, L64

\bibitem[\protect\citeauthoryear{{Eddington}}{{Eddington}}{1916}]{1916MNRAS..76..572E}
{Eddington} A.~S.,  1916, \mnras, 76, 572

\bibitem[\protect\citeauthoryear{{Feretti}, {Giovannini}, {Govoni} \&
  {Murgia}}{{Feretti} et~al.}{2012}]{2012A&ARv..20...54F}
{Feretti} L.,  {Giovannini} G.,  {Govoni} F.,    {Murgia} M.,  2012, \aapr, 20,
  54

\bibitem[\protect\citeauthoryear{{Gao}, {De Lucia}, {White} \& {Jenkins}}{{Gao}
  et~al.}{2004}]{2004MNRAS.352L...1G}
{Gao} L.,  {De Lucia} G.,  {White} S.~D.~M.,    {Jenkins} A.,  2004, \mnras,
  352, L1

\bibitem[\protect\citeauthoryear{{Giocoli}, {Bartelmann}, {Sheth} \&
  {Cacciato}}{{Giocoli} et~al.}{2010}]{2010MNRAS.408..300G}
{Giocoli} C.,  {Bartelmann} M.,  {Sheth} R.~K.,    {Cacciato} M.,  2010,
  \mnras, 408, 300

\bibitem[\protect\citeauthoryear{{Gonzalez}, {Sivanandam}, {Zabludoff} \&
  {Zaritsky}}{{Gonzalez} et~al.}{2013}]{2013ApJ...778...14G}
{Gonzalez} A.~H.,  {Sivanandam} S.,  {Zabludoff} A.~I.,    {Zaritsky} D.,
  2013, \apj, 778, 14

\bibitem[\protect\citeauthoryear{{Govoni} \& {Feretti}}{{Govoni} \&
  {Feretti}}{2004}]{2004IJMPD..13.1549G}
{Govoni} F.,  {Feretti} L.,  2004, International Journal of Modern Physics D,
  13, 1549

\bibitem[\protect\citeauthoryear{GSL~Project}{GSL~Project}{2010}]{contributors-gsl-gnu-2010}
GSL~Project C., , 2010, {GSL} - {GNU} Scientific Library - {GNU} Project - Free
  Software Foundation {(FSF)}, http://www.gnu.org/software/gsl/

\bibitem[\protect\citeauthoryear{{Gu}, {Xu}, {Gu}, {Kawaharada}, {Nakazawa},
  {Qin}, {Wang}, {Wang}, {Zhang} \& {Makishima}}{{Gu}
  et~al.}{2012}]{2012ApJ...749..186G}
{Gu} L.,  {Xu} H.,  {Gu} J.,  {Kawaharada} M.,  {Nakazawa} K.,  {Qin} Z.,
  {Wang} J.,  {Wang} Y.,  {Zhang} Z.,    {Makishima} K.,  2012, \apj, 749, 186

\bibitem[\protect\citeauthoryear{{Hernquist}}{{Hernquist}}{1990}]{1990ApJ...356..359H}
{Hernquist} L.,  1990, \apj, 356, 359

\bibitem[\protect\citeauthoryear{{Hoeft}, {Br{\"u}ggen}, {Yepes},
  {Gottl{\"o}ber} \& {Schwope}}{{Hoeft} et~al.}{2008}]{2008MNRAS.391.1511H}
{Hoeft} M.,  {Br{\"u}ggen} M.,  {Yepes} G.,  {Gottl{\"o}ber} S.,    {Schwope}
  A.,  2008, \mnras, 391, 1511

\bibitem[\protect\citeauthoryear{{Hong}, {Kang} \& {Ryu}}{{Hong}
  et~al.}{2015}]{2015ApJ...812...49H}
{Hong} S.~E.,  {Kang} H.,    {Ryu} D.,  2015, \apj, 812, 49

\bibitem[\protect\citeauthoryear{{Hong}, {Ryu}, {Kang} \& {Cen}}{{Hong}
  et~al.}{2014}]{2014ApJ...785..133H}
{Hong} S.~E.,  {Ryu} D.,  {Kang} H.,    {Cen} R.,  2014, \apj, 785, 133

\bibitem[\protect\citeauthoryear{{Iapichino} \& {Br{\"u}ggen}}{{Iapichino} \&
  {Br{\"u}ggen}}{2012}]{2012MNRAS.423.2781I}
{Iapichino} L.,  {Br{\"u}ggen} M.,  2012, \mnras, 423, 2781

\bibitem[\protect\citeauthoryear{{Jee}, {Stroe}, {Dawson}, {Wittman},
  {Hoekstra}, {Br{\"u}ggen}, {R{\"o}ttgering}, {Sobral} \& {van Weeren}}{{Jee}
  et~al.}{2015}]{2015ApJ...802...46J}
{Jee} M.~J.,  {Stroe} A.,  {Dawson} W.,  {Wittman} D.,  {Hoekstra} H.,
  {Br{\"u}ggen} M.,  {R{\"o}ttgering} H.,  {Sobral} D.,    {van Weeren} R.~J.,
  2015, \apj, 802, 46

\bibitem[\protect\citeauthoryear{{Kang} \& {Ryu}}{{Kang} \&
  {Ryu}}{2015}]{2015ApJ...809..186K}
{Kang} H.,  {Ryu} D.,  2015, \apj, 809, 186

\bibitem[\protect\citeauthoryear{{Kang} \& {Ryu}}{{Kang} \&
  {Ryu}}{2016}]{2016ApJ...823...13K}
{Kang} H.,  {Ryu} D.,  2016, \apj, 823, 13

\bibitem[\protect\citeauthoryear{{Kazantzidis}, {Magorrian} \&
  {Moore}}{{Kazantzidis} et~al.}{2004}]{2004ApJ...601...37K}
{Kazantzidis} S.,  {Magorrian} J.,    {Moore} B.,  2004, \apj, 601, 37

\bibitem[\protect\citeauthoryear{{Kocevski}, {Ebeling}, {Mullis} \&
  {Tully}}{{Kocevski} et~al.}{2007}]{2007ApJ...662..224K}
{Kocevski} D.~D.,  {Ebeling} H.,  {Mullis} C.~R.,    {Tully} R.~B.,  2007,
  \apj, 662, 224

\bibitem[\protect\citeauthoryear{{Kovesi}}{{Kovesi}}{2015}]{2015arXiv150903700K}
{Kovesi} P.,  2015, ArXiv e-prints

\bibitem[\protect\citeauthoryear{{Kravtsov} \& {Borgani}}{{Kravtsov} \&
  {Borgani}}{2012}]{2012ARA&A..50..353K}
{Kravtsov} A.~V.,  {Borgani} S.,  2012, \araa, 50, 353

\bibitem[\protect\citeauthoryear{{Lage} \& {Farrar}}{{Lage} \&
  {Farrar}}{2014}]{2014ApJ...787..144L}
{Lage} C.,  {Farrar} G.,  2014, \apj, 787, 144

\bibitem[\protect\citeauthoryear{{Markevitch}}{{Markevitch}}{2006}]{2006ESASP.604..723M}
{Markevitch} M.,  2006, in {Wilson} A.,  ed., The X-ray Universe 2005 Vol.~604
  of ESA Special Publication, {Chandra Observation of the Most Interesting
  Cluster in the Universe}.
p.~723

\bibitem[\protect\citeauthoryear{{Markevitch} \& {Vikhlinin}}{{Markevitch} \&
  {Vikhlinin}}{2007}]{2007PhR...443....1M}
{Markevitch} M.,  {Vikhlinin} A.,  2007, \physrep, 443, 1

\bibitem[\protect\citeauthoryear{{Mazzotta}, {Rasia}, {Moscardini} \&
  {Tormen}}{{Mazzotta} et~al.}{2004}]{2004MNRAS.354...10M}
{Mazzotta} P.,  {Rasia} E.,  {Moscardini} L.,    {Tormen} G.,  2004, \mnras,
  354, 10

\bibitem[\protect\citeauthoryear{{Miniati}, {Ryu}, {Kang}, {Jones}, {Cen} \&
  {Ostriker}}{{Miniati} et~al.}{2000}]{2000ApJ...542..608M}
{Miniati} F.,  {Ryu} D.,  {Kang} H.,  {Jones} T.~W.,  {Cen} R.,    {Ostriker}
  J.~P.,  2000, \apj, 542, 608

\bibitem[\protect\citeauthoryear{{Navarro}, {Frenk} \& {White}}{{Navarro}
  et~al.}{1996}]{1996ApJ...462..563N}
{Navarro} J.~F.,  {Frenk} C.~S.,    {White} S.~D.~M.,  1996, \apj, 462, 563

\bibitem[\protect\citeauthoryear{{Ogrean}, {Br{\"u}ggen}, {R{\"o}ttgering},
  {Simionescu}, {Croston}, {van Weeren} \& {Hoeft}}{{Ogrean}
  et~al.}{2013}]{2013MNRAS.429.2617O}
{Ogrean} G.~A.,  {Br{\"u}ggen} M.,  {R{\"o}ttgering} H.,  {Simionescu} A.,
  {Croston} J.~H.,  {van Weeren} R.,    {Hoeft} M.,  2013, \mnras, 429, 2617

\bibitem[\protect\citeauthoryear{{Ogrean}, {Br{\"u}ggen}, {van Weeren},
  {R{\"o}ttgering}, {Simionescu}, {Hoeft} \& {Croston}}{{Ogrean}
  et~al.}{2014}]{2014MNRAS.440.3416O}
{Ogrean} G.~A.,  {Br{\"u}ggen} M.,  {van Weeren} R.,  {R{\"o}ttgering} H.,
  {Simionescu} A.,  {Hoeft} M.,    {Croston} J.~H.,  2014, \mnras, 440, 3416

\bibitem[\protect\citeauthoryear{{Okabe}, {Akamatsu}, {Kakuwa}, {Fujita},
  {Zhang}, {Tanaka} \& {Umetsu}}{{Okabe} et~al.}{2015}]{2015PASJ...67..114O}
{Okabe} N.,  {Akamatsu} H.,  {Kakuwa} J.,  {Fujita} Y.,  {Zhang} Y.,  {Tanaka}
  M.,    {Umetsu} K.,  2015, \pasj, 67, 114

\bibitem[\protect\citeauthoryear{{Petrosian}}{{Petrosian}}{2001}]{2001ApJ...557..560P}
{Petrosian} V.,  2001, \apj, 557, 560

\bibitem[\protect\citeauthoryear{{Pfrommer}, {Springel}, {En{\ss}lin} \&
  {Jubelgas}}{{Pfrommer} et~al.}{2006}]{2006MNRAS.367..113P}
{Pfrommer} C.,  {Springel} V.,  {En{\ss}lin} T.~A.,    {Jubelgas} M.,  2006,
  \mnras, 367, 113

\bibitem[\protect\citeauthoryear{{Planck Collaboration}, {Ade}, {Aghanim},
  {Armitage-Caplan}, {Arnaud}, {Ashdown}, {Atrio-Barandela}, {Aumont},
  {Baccigalupi}, {Banday} \& et al.}{{Planck Collaboration}
  et~al.}{2014}]{2014A&A...571A..16P}
{Planck Collaboration} {Ade} P.~A.~R.,  {Aghanim} N.,  {Armitage-Caplan} C.,
  {Arnaud} M.,  {Ashdown} M.,  {Atrio-Barandela} F.,  {Aumont} J.,
  {Baccigalupi} C.,  {Banday} A.~J.,    et al. 2014, \aap, 571, A16

\bibitem[\protect\citeauthoryear{{Press}, {Teukolsky}, {Vetterling} \&
  {Flannery}}{{Press} et~al.}{1992}]{1992nrfa.book.....P}
{Press} W.~H.,  {Teukolsky} S.~A.,  {Vetterling} W.~T.,    {Flannery} B.~P.,
  1992, {Numerical recipes in FORTRAN. The art of scientific computing}.
Cambridge: University Press, |c1992, 2nd ed.

\bibitem[\protect\citeauthoryear{{Roettiger}, {Burns} \& {Stone}}{{Roettiger}
  et~al.}{1999}]{1999ApJ...518..603R}
{Roettiger} K.,  {Burns} J.~O.,    {Stone} J.~M.,  1999, \apj, 518, 603

\bibitem[\protect\citeauthoryear{{Russell}, {McNamara}, {Sanders}, {Fabian},
  {Nulsen}, {Canning}, {Baum}, {Donahue}, {Edge}, {King} \& {O'Dea}}{{Russell}
  et~al.}{2012}]{2012MNRAS.423..236R}
{Russell} H.~R.,  {McNamara} B.~R.,  {Sanders} J.~S.,  {Fabian} A.~C.,
  {Nulsen} P.~E.~J.,  {Canning} R.~E.~A.,  {Baum} S.~A.,  {Donahue} M.,  {Edge}
  A.~C.,  {King} L.~J.,    {O'Dea} C.~P.,  2012, \mnras, 423, 236

\bibitem[\protect\citeauthoryear{{Sarazin}}{{Sarazin}}{1988}]{1988xrec.book.....S}
{Sarazin} C.~L.,  1988, {X-ray emission from clusters of galaxies}

\bibitem[\protect\citeauthoryear{{Sarazin}}{{Sarazin}}{2002}]{2002ASSL..272....1S}
{Sarazin} C.~L.,  2002, in {Feretti} L.,  {Gioia} I.~M.,   {Giovannini} G.,
  eds, Merging Processes in Galaxy Clusters Vol.~272 of Astrophysics and Space
  Science Library, {The Physics of Cluster Mergers}.
pp 1--38

\bibitem[\protect\citeauthoryear{{Schaal}, {Springel}, {Pakmor}, {Pfrommer},
  {Nelson}, {Vogelsberger}, {Genel}, {Pillepich}, {Sijacki} \&
  {Hernquist}}{{Schaal} et~al.}{2016}]{2016MNRAS.461.4441S}
{Schaal} K.,  {Springel} V.,  {Pakmor} R.,  {Pfrommer} C.,  {Nelson} D.,
  {Vogelsberger} M.,  {Genel} S.,  {Pillepich} A.,  {Sijacki} D.,
  {Hernquist} L.,  2016, \mnras, 461, 4441

\bibitem[\protect\citeauthoryear{{Schindler} \& {Mueller}}{{Schindler} \&
  {Mueller}}{1993}]{1993A&A...272..137S}
{Schindler} S.,  {Mueller} E.,  1993, \aap, 272, 137

\bibitem[\protect\citeauthoryear{{Simionescu}, {Allen}, {Mantz}, {Werner},
  {Takei}, {Morris}, {Fabian}, {Sanders}, {Nulsen}, {George} \&
  {Taylor}}{{Simionescu} et~al.}{2011}]{2011Sci...331.1576S}
{Simionescu} A.,  {Allen} S.~W.,  {Mantz} A.,  {Werner} N.,  {Takei} Y.,
  {Morris} R.~G.,  {Fabian} A.~C.,  {Sanders} J.~S.,  {Nulsen} P.~E.~J.,
  {George} M.~R.,    {Taylor} G.~B.,  2011, Science, 331, 1576

\bibitem[\protect\citeauthoryear{{Skillman}, {O'Shea}, {Hallman}, {Burns} \&
  {Norman}}{{Skillman} et~al.}{2008}]{2008ApJ...689.1063S}
{Skillman} S.~W.,  {O'Shea} B.~W.,  {Hallman} E.~J.,  {Burns} J.~O.,
  {Norman} M.~L.,  2008, \apj, 689, 1063

\bibitem[\protect\citeauthoryear{{Sobral}, {Stroe}, {Dawson}, {Wittman}, {Jee},
  {R{\"o}ttgering}, {van Weeren} \& {Br{\"u}ggen}}{{Sobral}
  et~al.}{2015}]{2015MNRAS.450..630S}
{Sobral} D.,  {Stroe} A.,  {Dawson} W.~A.,  {Wittman} D.,  {Jee} M.~J.,
  {R{\"o}ttgering} H.,  {van Weeren} R.~J.,    {Br{\"u}ggen} M.,  2015, \mnras,
  450, 630

\bibitem[\protect\citeauthoryear{{Spitzer}}{{Spitzer}}{1956}]{1956pfig.book.....S}
{Spitzer} L.,  1956, {Physics of Fully Ionized Gases}

\bibitem[\protect\citeauthoryear{{Stroe}, {Harwood}, {Hardcastle} \&
  {R{\"o}ttgering}}{{Stroe} et~al.}{2014}]{2014MNRAS.445.1213S}
{Stroe} A.,  {Harwood} J.~J.,  {Hardcastle} M.~J.,    {R{\"o}ttgering}
  H.~J.~A.,  2014, \mnras, 445, 1213

\bibitem[\protect\citeauthoryear{{Stroe}, {Rumsey}, {Harwood}, {van Weeren},
  {R{\"o}ttgering}, {Saunders}, {Sobral}, {Perrott} \& {Schammel}}{{Stroe}
  et~al.}{2014}]{2014MNRAS.441L..41S}
{Stroe} A.,  {Rumsey} C.,  {Harwood} J.~J.,  {van Weeren} R.~J.,
  {R{\"o}ttgering} H.~J.~A.,  {Saunders} R.~D.~E.,  {Sobral} D.,  {Perrott}
  Y.~C.,    {Schammel} M.~P.,  2014, \mnras, 441, L41

\bibitem[\protect\citeauthoryear{{Stroe}, {Shimwell}, {Rumsey}, {van Weeren},
  {Kierdorf}, {Donnert}, {Jones}, {R{\"o}ttgering}, {Hoeft},
  {Rodr{\'{\i}}guez-Gonz{\'a}lvez}, {Harwood} \& {Saunders}}{{Stroe}
  et~al.}{2016}]{2016MNRAS.455.2402S}
{Stroe} A.,  {Shimwell} T.,  {Rumsey} C.,  {van Weeren} R.,  {Kierdorf} M.,
  {Donnert} J.,  {Jones} T.~W.,  {R{\"o}ttgering} H.~J.~A.,  {Hoeft} M.,
  {Rodr{\'{\i}}guez-Gonz{\'a}lvez} C.,  {Harwood} J.~J.,    {Saunders}
  R.~D.~E.,  2016, \mnras, 455, 2402

\bibitem[\protect\citeauthoryear{{Stroe}, {Sobral}, {Dawson}, {Jee},
  {Hoekstra}, {Wittman}, {van Weeren}, {Br{\"u}ggen} \&
  {R{\"o}ttgering}}{{Stroe} et~al.}{2015}]{2015MNRAS.450..646S}
{Stroe} A.,  {Sobral} D.,  {Dawson} W.,  {Jee} M.~J.,  {Hoekstra} H.,
  {Wittman} D.,  {van Weeren} R.~J.,  {Br{\"u}ggen} M.,    {R{\"o}ttgering}
  H.~J.~A.,  2015, \mnras, 450, 646

\bibitem[\protect\citeauthoryear{{Stroe}, {Sobral}, {R{\"o}ttgering} \& {van
  Weeren}}{{Stroe} et~al.}{2014}]{2014MNRAS.438.1377S}
{Stroe} A.,  {Sobral} D.,  {R{\"o}ttgering} H.~J.~A.,    {van Weeren} R.~J.,
  2014, \mnras, 438, 1377

\bibitem[\protect\citeauthoryear{{Stroe}, {van Weeren}, {Intema},
  {R{\"o}ttgering}, {Br{\"u}ggen} \& {Hoeft}}{{Stroe}
  et~al.}{2013}]{2013AA...555A.110S}
{Stroe} A.,  {van Weeren} R.~J.,  {Intema} H.~T.,  {R{\"o}ttgering} H.~J.~A.,
  {Br{\"u}ggen} M.,    {Hoeft} M.,  2013, \aap, 555, A110

\bibitem[\protect\citeauthoryear{{Tormen}, {Diaferio} \& {Syer}}{{Tormen}
  et~al.}{1998}]{1998MNRAS.299..728T}
{Tormen} G.,  {Diaferio} A.,    {Syer} D.,  1998, \mnras, 299, 728

\bibitem[\protect\citeauthoryear{{Urban}, {Simionescu}, {Werner}, {Allen},
  {Ehlert}, {Zhuravleva}, {Morris}, {Fabian}, {Mantz}, {Nulsen}, {Sanders} \&
  {Takei}}{{Urban} et~al.}{2014}]{2014MNRAS.437.3939U}
{Urban} O.,  {Simionescu} A.,  {Werner} N.,  {Allen} S.~W.,  {Ehlert} S.,
  {Zhuravleva} I.,  {Morris} R.~G.,  {Fabian} A.~C.,  {Mantz} A.,  {Nulsen}
  P.~E.~J.,  {Sanders} J.~S.,    {Takei} Y.,  2014, \mnras, 437, 3939

\bibitem[\protect\citeauthoryear{{van Weeren}, {Br{\"u}ggen}, {R{\"o}ttgering}
  \& {Hoeft}}{{van Weeren} et~al.}{2011}]{2011MNRAS.418..230V}
{van Weeren} R.~J.,  {Br{\"u}ggen} M.,  {R{\"o}ttgering} H.~J.~A.,    {Hoeft}
  M.,  2011, \mnras, 418, 230

\bibitem[\protect\citeauthoryear{{van Weeren}, {R{\"o}ttgering}, {Br{\"u}ggen}
  \& {Hoeft}}{{van Weeren} et~al.}{2010}]{2010Sci...330..347V}
{van Weeren} R.~J.,  {R{\"o}ttgering} H.~J.~A.,  {Br{\"u}ggen} M.,    {Hoeft}
  M.,  2010, Science, 330, 347

\bibitem[\protect\citeauthoryear{{Vazza}, {Brunetti} \& {Gheller}}{{Vazza}
  et~al.}{2009}]{2009MNRAS.395.1333V}
{Vazza} F.,  {Brunetti} G.,    {Gheller} C.,  2009, \mnras, 395, 1333

\bibitem[\protect\citeauthoryear{{Vikhlinin}, {Kravtsov}, {Forman}, {Jones},
  {Markevitch}, {Murray} \& {Van Speybroeck}}{{Vikhlinin}
  et~al.}{2006}]{2006ApJ...640..691V}
{Vikhlinin} A.,  {Kravtsov} A.,  {Forman} W.,  {Jones} C.,  {Markevitch} M.,
  {Murray} S.~S.,    {Van Speybroeck} L.,  2006, \apj, 640, 691

\bibitem[\protect\citeauthoryear{{Zhuravleva}, {Churazov}, {Sunyaev},
  {Sazonov}, {Allen}, {Werner}, {Simionescu}, {Konami} \&
  {Ohashi}}{{Zhuravleva} et~al.}{2013}]{2013MNRAS.435.3111Z}
{Zhuravleva} I.,  {Churazov} E.,  {Sunyaev} R.,  {Sazonov} S.,  {Allen} S.~W.,
  {Werner} N.,  {Simionescu} A.,  {Konami} S.,    {Ohashi} T.,  2013, \mnras,
  435, 3111

\bibitem[\protect\citeauthoryear{{ZuHone} \& {Kowalik}}{{ZuHone} \&
  {Kowalik}}{2016}]{2016arXiv160904121Z}
{ZuHone} J.~A.,  {Kowalik} K.,  2016, ArXiv e-prints

\end{thebibliography}

\label{lastpage}
\end{document}